\documentclass[aps,reprint,prl,groupedaddress]{revtex4-2}
\usepackage{xcolor}
\usepackage{float}
\usepackage{amsmath,amsfonts,amssymb,epsfig,graphicx}
\usepackage{slashed} 
\usepackage{hyperref}
\usepackage[normalem]{ulem}
\usepackage{lineno}
\graphicspath{{./pic/}}
\begin{document}

\title{A proposed PKU-Muon experiment for muon tomography and dark matter search}

\author{Xudong \surname{Yu}}
\email[]{yuxd@stu.pku.edu.cn }
\affiliation{State Key Laboratory of Nuclear Physics and Technology, School of Physics, Peking University, Beijing, 100871, China}

\author{Zijian \surname{Wang}}
%\email[]{ wangzijian@stu.pku.edu.cn}
\affiliation{State Key Laboratory of Nuclear Physics and Technology, School of Physics, Peking University, Beijing, 100871, China}

\author{Cheng-en \surname{Liu}}
%\email[]{2301210069@stu.pku.edu.cn }
\affiliation{State Key Laboratory of Nuclear Physics and Technology, School of Physics, Peking University, Beijing, 100871, China}

\author{Yiqing \surname{Feng}}
%\email[]{2000011324@stu.pku.edu.cn }
\affiliation{State Key Laboratory of Nuclear Physics and Technology, School of Physics, Peking University, Beijing, 100871, China}

\author{Jinning \surname{Li}}
%\email[]{lijinning@stu.pku.edu.cn}
\affiliation{State Key Laboratory of Nuclear Physics and Technology, School of Physics, Peking University, Beijing, 100871, China}

\author{Xinyue \surname{Geng}}
%\email[]{2100011516@stu.pku.edu.cn }
\affiliation{State Key Laboratory of Nuclear Physics and Technology, School of Physics, Peking University, Beijing, 100871, China}

\author{Yimeng \surname{Zhang}}
%\email[]{2100011516@stu.pku.edu.cn }
\affiliation{State Key Laboratory of Nuclear Physics and Technology, School of Physics, Peking University, Beijing, 100871, China}

\author{Leyun \surname{Gao}}
%\email[]{seeson@pku.edu.cn }
\affiliation{State Key Laboratory of Nuclear Physics and Technology, School of Physics, Peking University, Beijing, 100871, China}

\author{Ruobing \surname{Jiang}}
%\email[]{ruobingjiang2019@outlook.com }
\affiliation{State Key Laboratory of Nuclear Physics and Technology, School of Physics, Peking University, Beijing, 100871, China}

\author{Youpeng \surname{Wu}}
%\email[]{wuyoupeng2001@outlook.com}
\affiliation{State Key Laboratory of Nuclear Physics and Technology, School of Physics, Peking University, Beijing, 100871, China}

\author{Chen \surname{Zhou}}
\email[]{czhouphy@pku.edu.cn}
\affiliation{State Key Laboratory of Nuclear Physics and Technology, School of Physics, Peking University, Beijing, 100871, China}

\author{Qite \surname{Li}}
\email[]{liqt@pku.edu.cn}
\affiliation{State Key Laboratory of Nuclear Physics and Technology, School of Physics, Peking University, Beijing, 100871, China}

\author{Siguang \surname{Wang}}
%\email[]{siguang@pku.edu.cn}
\affiliation{State Key Laboratory of Nuclear Physics and Technology, School of Physics, Peking University, Beijing, 100871, China}

\author{Yong \surname{Ban}}
%\email[]{bany@pku.edu.cn}
\affiliation{State Key Laboratory of Nuclear Physics and Technology, School of Physics, Peking University, Beijing, 100871, China}

\author{Yajun \surname{Mao}}
%\email[]{maoyj@pku.edu.cn}
\affiliation{State Key Laboratory of Nuclear Physics and Technology, School of Physics, Peking University, Beijing, 100871, China}

\author{Qiang \surname{Li}}
\email[]{qliphy0@pku.edu.cn}
\affiliation{State Key Laboratory of Nuclear Physics and Technology, School of Physics, Peking University, Beijing, 100871, China}

\begin{abstract}
We propose here a set of new methods to directly detect light mass dark matter through its scattering with abundant atmospheric muons or accelerator beams. Firstly, we plan to use the free cosmic-ray muons interacting with dark matter in a volume surrounded by tracking detectors, to trace possible interaction between dark matter and muons. Secondly, we will interface our device with domestic or international muon beams. Due to much larger muon intensity and focused beam, we anticipate the detector can be made further compact and the resulting sensitivity on dark matter searches will be improved. %In line with above projects, we will develop muon tomography methods and apply them on atmospheric and environmental sciences, archaeology and civil engineering. 
Furthermore, we will measure precisely directional distributions of cosmic-ray muons, either at mountain or sea level, and the differences may reveal possible information of dark matter distributed near the earth. %In the future, we may also extend our study to muon on target experiments. 
Specifically, our methods can have advantages over `exotic' dark matters which are either muon-philic or slowed down due to some mechanism, and sensitivity on dark matter and muon scattering cross section can reach as low as microbarn level.
\end{abstract}

\maketitle

\section{Introduction and Motivations} 

Among all the elementary particles, muon has its speciality to connect applied studies with fundamental researches, yet both to a lesser extent compared with its cousin, $e.g.$, electron.

Firstly, using cosmic-ray muons to image the internal structure of objects, especially large-scale objects, has become a booming research area in particle physics. The earliest application of muography can be traced back to 1955, when George from Australia measured overburden of tunnel~\cite{george1955cosmic}. In the following decades, muography has been widely used in the field of geology, including the exploration of rock formations~\cite{1979Theoretical,2017Muography}, glaciers~\cite{2017First,nishiyama2019bedrock}, minerals~\cite{bonneville2019borehole,zhang2020muography}, oceans~\cite{tanaka2021first} and underground carbon dioxide storage~\cite{kudryavtsev2012monitoring,klinger2015simulation,gluyas2019passive}, $etc$.

In 1969, a team led by Alvarez pioneered muography by measuring the internal structures of the Giza and Chephren pyramids in Egypt~
\cite{alvarez1970search}. This marked the first application of muography in archaeology, showcasing its ability to non-destructively reveal the hidden features of ancient ruins. Since the 21st century, muography has played a crucial role in discovering hidden chambers within structures like the Pyramid of the Sun in Teotihuacan~\cite{aguilar2013using} and Khufu’s Pyramid~\cite{morishima2017discovery}. In 2021, researchers from Beijing Normal University employed muography to measure the Mausoleum of Qin Shihuang, successfully reconstructing the size and location of the underground palace tomb~\cite{suning2020simulation}. In addition, muography has also aided the study of ancient earthquakes~\cite{anceart}.

In recent years, monitoring volcanoes has emerged as a prominent research area in muography. Numerous Japanese volcanoes, including Showa-Shinzan~\cite{tanaka2007imaging}, Asama~\cite{tanaka2009detecting}, Sakurajima~\cite{olah2018high}, and Stromboli in Italy~\cite{tioukov2019first}, has been subject to muographic studies. Similar investigations have been condected in France~\cite{ambrosino2015joint} and the United States~\cite{guerrero2019design}. In addition to large natural geological structures, there is a growing interest in combining muography with meteorology, exemplified by its application in monitoring tropical cyclones~\cite{tanaka2022atmospheric}.

Due to its non-destructive nature and safety, muon imaging presents a viable strategy for nuclear safety monitoring and detecting radioactive materials. Applications include the visualization of reactor interiors~\cite{takamatsu2015cosmic}, detection of spent nuclear fuel in dry storage barrels~\cite{poulson2017cosmic} and nuclear waste~\cite{mahon2019first}. However, the effectiveness of both muon radiography and muon tomography relies heavily on precise measurements of energy spectrum and angular distribution of cosmic-ray muons. Consequently, obtaining accurate data on muon flux is essential. Various sea-level muon flux models have been proposed and measured using diverse experimental methods in different geographical locations~\cite{autran2018characterization,su2021comparison,abbrescia2023measurement,conte2023brief}. Building on this foundation, some research teams have delved into understanding the impact of factors such as temperature, barometric pressure~\cite{alekseev2022observation}, and seasonal variations~\cite{an2018seasonal} on muon flux.

Cosmic-ray muons are characterized by low density with a mean energy of 3-4~GeV at sea level and a mean flux of approximately 1~cm$^{-2}$min$^{-1}$. To produce more intense and controllable muon beams, several countries have established their own muon sources.

For instance, the muon source at Paul Scherrer Institute (PSI) in Switzerland, S\textmu S, uses a continuous-wave high-current proton beam generated by two cascaded high-current proton cyclotrons. This beam is directed successively onto targets, providing secondary $\mu$ or $\pi$ beams, dedicated to various experiments, such as Mu3E, MUSE, \textmu SR~\cite{morenzoni2000low}, MuCOOL, MuSUN, HyperMu, MEG and Mu3E~\cite{2016The}.
%, namely: PiM1, PiM3, PiE1, PiE3, PiE5, MuE1, and MuE4. These seven beams are dedicated to various experiments. For example, PiM1 conducts particle physics experiments such as Mu3E and MUSE, PiE1 conducts experiments such as \textmu SR~\cite{morenzoni2000low}, MuCOOL, MuSUN, and HyperMu, while PiE5 is mainly used for experiments such as MEG and Mu3E~\cite{2016The}.
TRIUMF’s muon source CMMS has three muon beams, namely M9, M15, and M20 along with two series-connected muon targets, T1 and T2~\cite{triumf}. M9A, M15, and M20 are employed for conducting \textmu SR experiments primarily based on surface muon beams. M9B is dedicated to \textmu SR experiments using decaying muon beam with thick samples. Additionally, M11 is mainly used for detector testing. The ISIS muon source, located at the Rutherford Appleton Laboratory in the UK, is the world's first pulsed muon source~\cite{isis}. Equipped with a thin graphite muon target, it has two muon beams flanking the target. These beams are dedicated to conducting \textmu SR experiments, leveraging both surface muon beams and decay muon beams~\cite{hillier2014developments,2002A}. J-PARC in Japan, 
%jointly invested and constructed by KEK and JAEA,
%High Energy Accelerator Research Organization (KEK) and Japan Atomic Energy Agency (JAEA), 
houses two pulsed muon sources: MUSE, dedicated to \textmu SR experiments, and COMET, specializing in particle physics experiments~\cite{jparc}. 
%The design of MUSE is relatively complex~\cite{{miyake2010j,pant2019characterization,kawamura2013h}}, featuring an X-shaped muon beam configuration. There are 4 beams in total: H-line, D-line, S-line and U-line. Each muon beam is divided into multiple branch beams. Meanwhile, COMET focuses on exploring the rare decays of muons into electrons~\cite{litchfield2014muon}.
FNAL in the US is actively engaged in groundbreaking experiments, notably the g-2 experiment aimed at measuring the anomalous magnetic moment of muons~\cite{fermi,2017The}, and the Mu2e experiment, which focuses on investigating rare decays of muons into electrons at its muon campus~\cite{2015Mu2e,2014Muon}. Moreover, China and South Korea are also in the process of designing and planning the construction of their own new muon sources.

Both cosmic-ray and man-made muons stand out as powerful tools for delving into the mysteries of our world and the unknown. For example, the Standard Model (SM) of particle physics, while remarkably successful, falls short in explaining certain experimental observations, such as the neutrino mass and Dark Matter (DM). To address these gaps, physics beyond the SM is imperative, introducing hypothetical particles and novel interactions. Among these hypothesized particles, understanding the nature of DM is currently a focal point for both cosmology and particle physics. Despite our confidence in its existence, deduced from cosmological observations, no experiments have directly detected any form of DM particles.

There are several experiments dedicated to the direct or indirect detection of light DM particles, whose mass ranges from sub-MeV to GeV scales. The paradigm of Weakly Interacting Massive Particles (WIMPs) is extensively studied in the DM sector, with numerous experiments conducted to search for them. However, no observations of WIMPs have materialized, prompting the need to explore other theoretically motivated scenarios. Examples include low-mass or muon-philic DM~\cite{Essig:2022dfa,Harris:2022vnx,Bai:2014osa}.

%Detection sub-GeV DM is a more difficult challenge, and current limits in this region are much weaker than those for GeV-scale DM. 
Traditional direct searches for DM, involving the identification of nuclear recoils in deep underground detectors, are challenging due to insufficient recoil energy in the sub-MeV range. Experimental searches of cold sub-GeV DM have shifted focus towards exploring the Migdal effect~\cite{migdal} and interactions with electrons~\cite{EDELWEISS,SENSEI}. Proposals for low-mass DM searches using liquid helium have also been put forward~\cite{Liao:2021npo,Liao:2022zqg}. Recently, experiments also exploit the fact that a fraction of the cold DM is boosted to relativistic energies and thus can be efficiently detected in direct detection experiments~\cite{Super-Kamiokande:2022ncz,PandaX-II:2021kai,Bringmann:2018cvk,Plestid:2020kdm,Hu:2016xas}. More projections on low-threshold DM direct detection in the next decade can be found in Ref.~\cite{Essig:2022dfa,Elor:2021swj}.

Here we extend our previous study~\cite{Ruzi:2023mxp} and propose a set of new methods to directly detect light mass DM through its scattering with abundant atmospheric muons or accelerator beams, to directly probe muon-philic DM in a model-independent way.

Our methods can have also advantages over `exotic' DMs~\cite{McKeen:2023ztq} which can be well slowed down through scattering with matter in the atmosphere or the Earth before reaching the detector target. As shown below, methods here rely on high speed incoming particles and thus depend less on DM velocity. 

\section{Our proposal in brief} 

We are interested in exploiting muon detector such as RPC and GEM for both Muon tomography and dark matter searches. 

\begin{figure}
    \centering
    \includegraphics[width=1.\columnwidth]{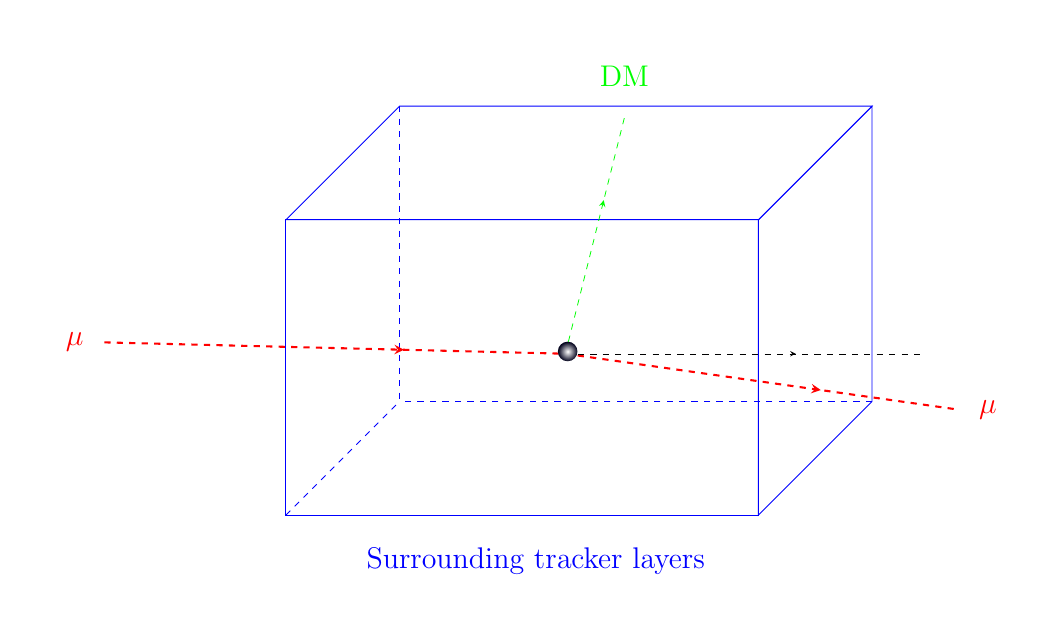}
    \caption{Illustration of an experiment to detect muon-philic DM with free leptons. The resulting kinematic shifts of leptons kicked by DM can be measurable with tracking detectors surrounding a vacuum region. A veto region along the chamber can be defined based on the cross-point of the in and out tracks to suppress backgrounds. }
    \label{fig:muonbox}
\end{figure}

As a first step as shown in Fig.~\ref{fig:muonbox}, we plan to use the free cosmic-ray muons interacting with dark matter in a volume surrounded by tracking detectors, to trace possible interaction between dark matter and muons. 

\begin{figure}
    \centering
    \includegraphics[width=1.\columnwidth]{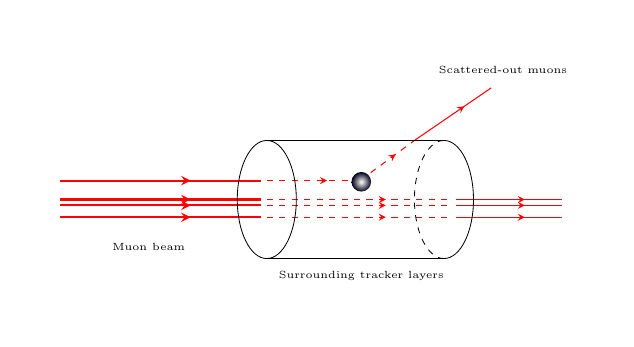}
    \caption{Illustration of an experiment to detect muon-philic DM with muon beams. Due to focused sizes of muon beams, the cylindrical detector can be made very compact.}
    \label{fig:mubeambox}
\end{figure}

We will then interface our device with domestic or international muon beams as shown in Fig.~\ref{fig:mubeambox}. Due to much larger muon intensity and focused beam, we anticipate the detector can be made further compact and the resulting sensitivity on dark matter searches will be further improved.

In line with above projects, we will develop muon tomography methods on both detectors and algorithm parts. Depending on possible precision to be achieved, we may also apply muon tomography on atmospheric and environmental sciences, on archaeology and civil engineering. 

\begin{figure}
    \centering
    \includegraphics[width=1.\columnwidth]{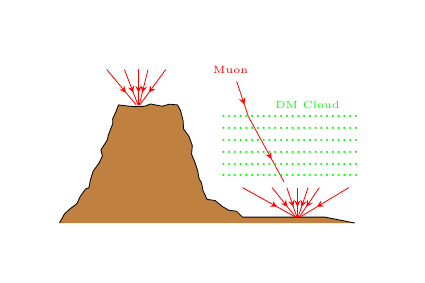}
    \caption{Illustration of an experiment to detect muon-philic DM through precisely measuring directional distributions of cosmic-ray muons, either at mountain or sea level.}
    \label{fig:moutain}
\end{figure}

Furthermore, we will measure precisely directional distributions of cosmic-ray muons, either at mountain or sea level as shown in Fig.~\ref{fig:moutain}, and the differences may reveal possible information of dark matter distributed near the earth. 

\section{Experiment Setup: GEM and RPC} 
\label{sec:device}
\begin{figure}
    \centering
    \includegraphics[width=1.\columnwidth]{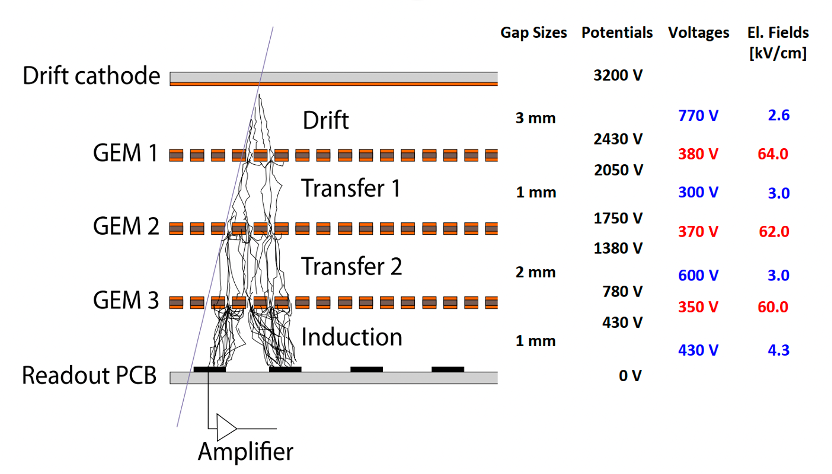}
    \caption{Schematic view of a triple-GEM detector~\cite{CMS:2017gemtdr}.}
    \label{fig:GEMPicture}
\end{figure}

In order to meet the requirements of muon tomography and dark matter searches as describe above, the track detectors must contain a list of conditions: large area ($\sim$ 1~m$^{2}$), high detection efficiency ($>90\%$), high spatial resolution ($\sim$ 1~mm), high timing resolution ($\sim$ 1~ns) and low cost. 

Gas Electron Multiplier (GEM)~\cite{Sauli:1997qp} detectors are widely used for high-energy physics experiments, such as the triple-GEM detector~\cite{CMS:2017gemtdr} installed in the CMS experiment, due to their excellent performance.
%The GEM (Gas Electron Multiplier) technology has been adopted by many particle physics experiments, such as the high-luminosity upgrade of the CMS detector. 
The GEM detectors can improve trigger capabilities and muon measurements in the forward region of CMS because of their excellent performance: a rate capability above 10~kHZ/cm$^2$, a time resolution of about 8~ns and a spatial resolution of about 200~$\mu$m.

A GEM foil is a metal-clad polyimide foil which is perforated by a high density of holes. A triple-GEM detector~\cite{CMS:2017gemtdr} consists of three GEM foils which are immersed in a gas mixture and placed between a cathode drift board and an anode readout board. The drift board routes a total of seven potentials (drift electrode plus both faces of the three GEM foils) while the readout board is grounded. A schematic view of a triple-GEM detector is shown in Figure~\ref{fig:GEMPicture}. 

A charged particle passing the triple-GEM detector produces electrons due to ionization. The electrons experience intense electric field when moving through the GEM holes and produce secondary ionization. This electron avalanche process induces electric signals on the strips or pixels of the readout board. The signals from the readout board are processed and transmitted by front-end readout electronics and data acquisition system.

GEM detectors are distinguished by the use of GEM foils as the electron amplification structure, where the electron avalanche process is independent of the readout anode. This flexibility allows GEM detectors to adopt various readout structures.  However, since the signal in GEM detectors has minimal transverse diffusion of the signal (1-3 mm), in order to obtain a good space resolution, one has to reduce the size of the pixel and employ a large amount of pixels to cover a required effective area, which will lead to a great pressure on the detector construction, power consumption, spatial utilization, etc.

The resistive anode readout method~\cite{juxudong} can help to obtain a good spatial resolution comparable to the pixel readout structure with an enormous reduction of the electronic channels. By using the thick film resistor technology, a new type of resistive structure, composed of high resistive square pad array with low resistive narrow border strips, is developed and applied to the readout anode of the triple GEM detector. On the basis of that, We aim to design our exclusive readout for the specific requirements of PKU-Muon GEM detectors. 

\begin{figure}
    \centering
    \includegraphics[width=1.\columnwidth]{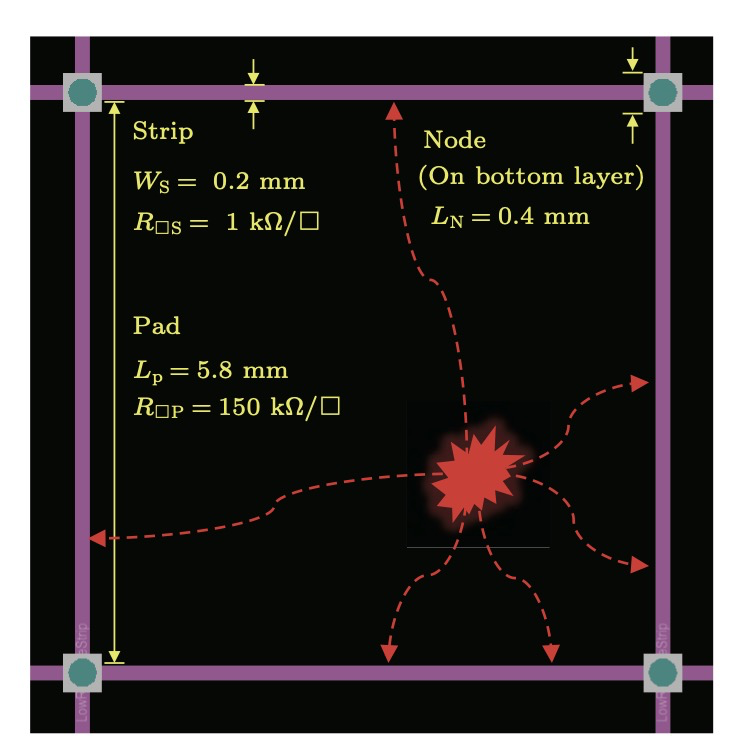}
    \caption{Structure diagram of the basic resistive anode cell.}
    \label{fig:GEM}
\end{figure}

Unlike commonly used conductive readout electrodes, where the induced charge is promptly collected, in resistive structures, there is a continuous diffusion process for charge induction. Charges will be collected by the readout nodes of the adjacent cells. Using the information of the collected charges, the hit position can be reconstructed with suitable algorithm. As shown in Fig.~\ref{fig:GEM}, the black square region represents a high-resistance pad, while the purple narrow line region represents a low-resistance strip. The gray small square (located on the backside) serves as the signal readout electrode. The induced charge will diffuse slowly on high-resistance pad. When the charge diffuses into the low-resistance region, it will preferentially spread along the low-resistance strip towards the readout electrode, as the resistance in this area is significantly lower than that in the adjacent high-resistance blocks. So we can calculate the hit position according to the quantity of electric charge that 4 readout electrode collected. Using this method, at the same time as maintaining high position resolution we can significantly reduce the required number of electronic channels. We will carry out further research and practical applications.

%Based on local expertise and the cost-effectiveness of the performance, we plan to exploit Gas Electron Multiplier (GEM) technology for muon tracking. GEM is a proven amplification technique for position detection of charged particles in gaseous detectors. GEM has been widely used in various high energy physics experiments, such as the CMS experiment~\cite{Abbas:2022fze,Pellecchia:2022lsd}. We plan to adopt the triple-GEM design from the CMS phase-II detector upgrades: in each chamber, three GEM foils with microscopic holes are stacked between an anode readout board and a cathode drift board. These detectors have shown good performances including detection efficiency, time resolution and spatial resolution. With multiple triple-GEM chambers surrounding sides of a vacuum cube, we aim at a position resolution around 100 microns~\cite{Abbas:2022fze,Pellecchia:2022lsd}. 

%The cost of our detector system is mainly driven by the front-end electronics and data-acquisition (DAQ) system. To provide spatial measurements for muons crossing a vacuum cube with a length of $L\sim 1$ meter, each triple-GEM chamber surrounded the cube sides needs 6400 readout channels, which cost 5,000,000 RMB. Hence, the total cost of front-end electronics and DAQ system for six triple-GEM chambers is approximately 30,000,000  RMB when working with a $L\sim 1$ meter vacuum cube. The price can be reduced significantly with a smaller cube (notice though the fiducial region and thus sensitivity will get degraded), or a cube with electronics covered only in the top and bottom sides, as a startup prototype.

\begin{figure}
    \centering
    \includegraphics[width=1.\columnwidth]{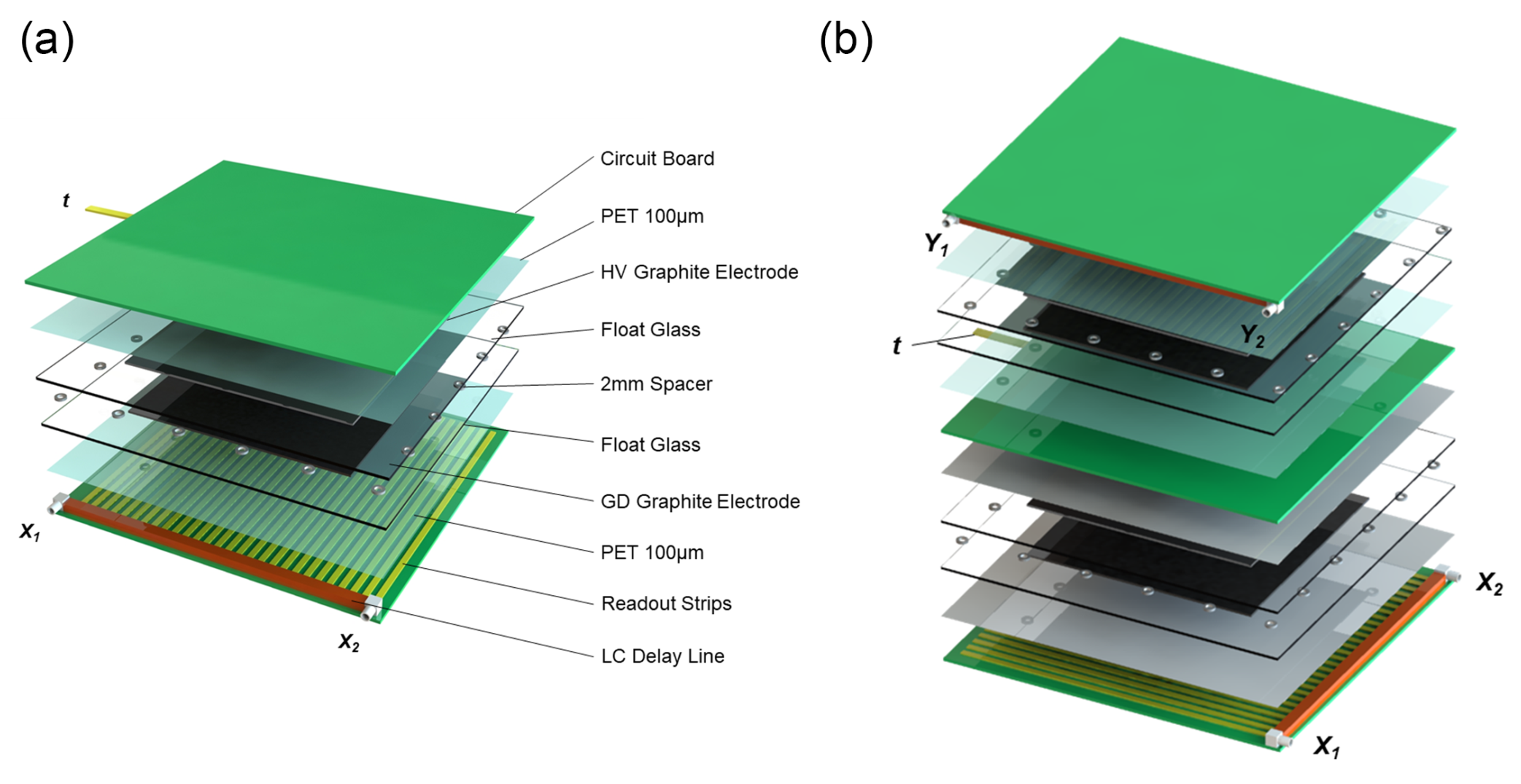}
    \caption{(a) Layout of the prototype glass RPC. (b) One RPC layer consists of two same structures, which get X and Y signals respectively. In addition, each layer has a timing signal channel t not passing the delay-line.}
    \label{fig:RPC}
\end{figure}

Resistive plate chambers (RPCs) have also been used widely in high energy physics field as particle detectors for nearly 50 years due to their advantages such as a simple and robust structure, long-term stability, good timing resolution, easy-maintenance and low cost. %Therefore, RPCs can well satisfy the demands and have been used to construct muon tomography systems.
Peking University is an early researcher in China on muon tomography by successfully building up a prototype in 2012, i.e., a 2-dimensional glass RPC detector with a large area based on LC delay-lines readout. The detailed design and assembly of the prototype glass RPC were described in Ref.~\cite{LI201222}. As shown in Fig.~\ref{fig:RPC}, two 2.6~mm thick float glass plates were laid in parallel with each area of $30 \times 30~{\rm cm}^{2}$. Using several 2~mm thick gaskets, the gas gap of 2~mm between two glass plates was set up. The graphite electrodes were coated on the glass plates to supply high voltage. The effective area of the electrode was $20 \times 20~{\rm cm}^{2}$. Top and bottom were insulated with $100~\mu{\rm m}$ PET layers. And then the outer layer was the printed-circuit board (PCB), with 2.54~mm readout strips on it. To save the electronics, we follow the delay-line, charge-division methods, which have been proved effective in Refs.~\cite{Li:2013PHC, Chen:JINST} and help to improve the spatial resolution as well. One end of each strip was connected to delay-lines with 4~ns delay between adjacent two strips. Signals from one end of delay-lines were fed into 300 MHz fast preamplifiers. To get X and Y signals in the same time, we put two prototypes in one aluminum box orthogonally, as shown in Fig.~\ref{fig:RPC}(b). In addition, each layer generates a timing signal channel t, which does not pass the delay-line. The box was then equipped with HV, signal and gas in-out connectors. Our all RPCs used here were constructed in the same way, and were stacked with an equal interval. It is able to decrease the positional resolution of cosmic-ray ray muons to submillimeter, while maintaining a detection efficiency above $90\%$.

In the near future, we plan to improve the performance of RPC detector. We will use mixed gas $(87\% \,\text{Freon} + 8.7\%\, \mathrm{C}_4\mathrm{H}_{10} + 4.3\%\, \mathrm{SF}_6)$ instead of pure Freon as working gas, so that avalanche mode dominated and in the meanwhile signal to noise ration kept a high level. Besides, to simplify the electronic readout, we will adopt a DAQ system based on digitization ASIC, such as AGET, PETIROC, TOFPET, $etc$.

\begin{figure}
    \centering
    \includegraphics[width=1.\columnwidth]{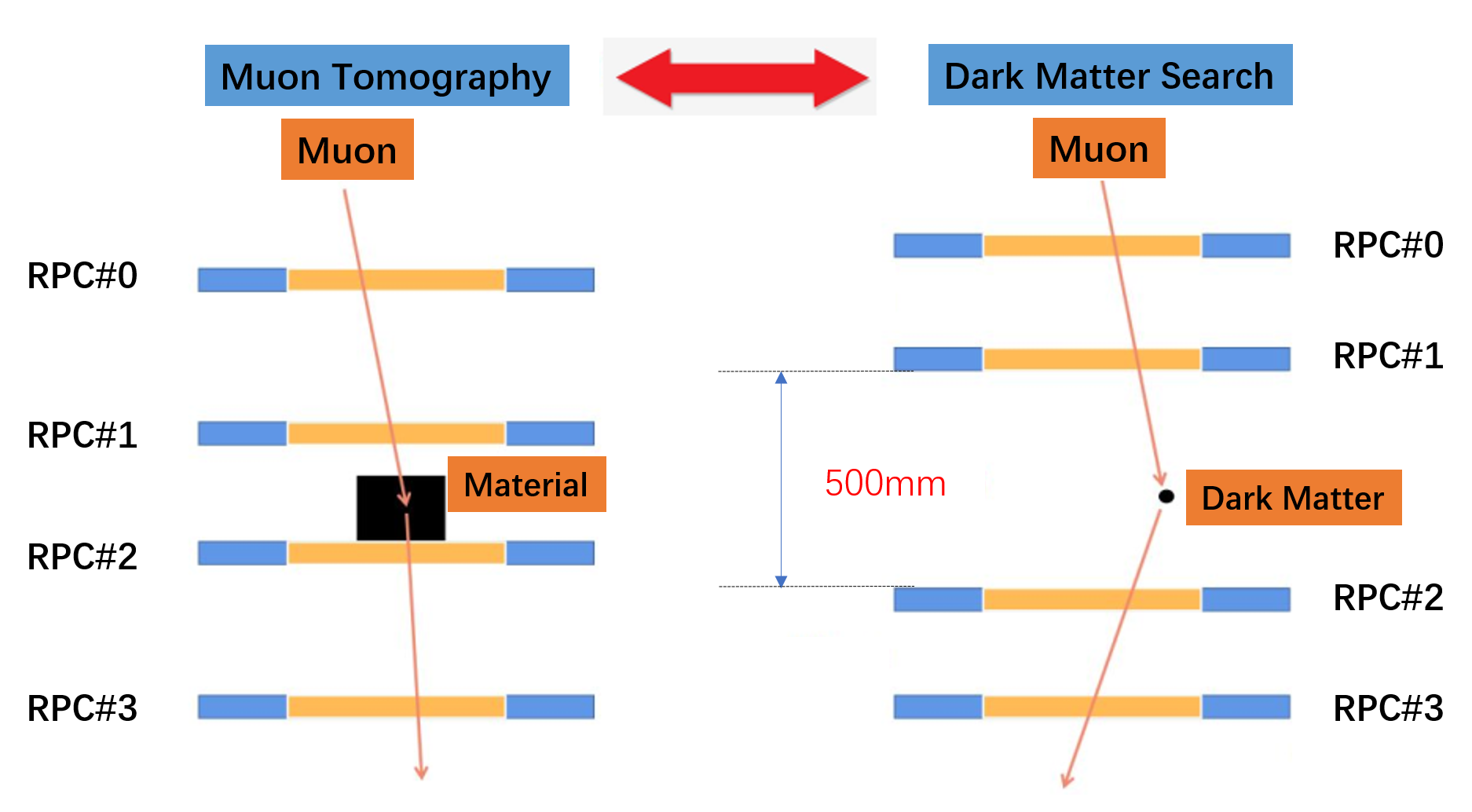}
    \caption{Schematic diagram of RPC testing system}
    \label{fig:schematic}
\end{figure}

On top of above mentioned mugraphy system, we will upgrade it to the mode of dark matter searching, as shown in Fig.~\ref{fig:schematic}.

\begin{figure}
    \centering
    \includegraphics[width=1.\columnwidth]{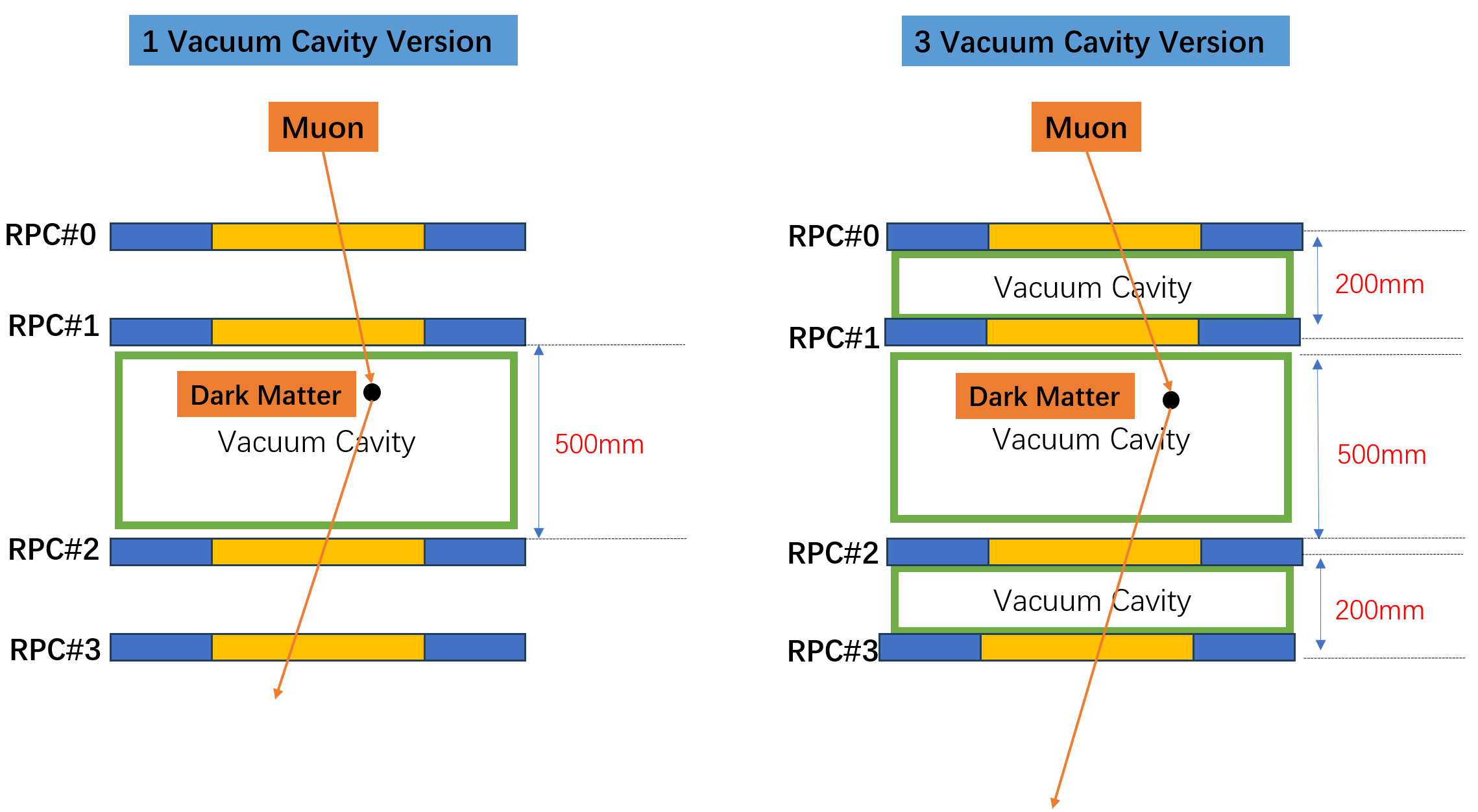}
    \caption{Schematic diagram of RPC testing system with vacuum cavity}
    \label{fig:vacuum}
\end{figure}

In order to minimize other factors interfering with our dark matter search, we plan to add a chamber in the middle of RPCs to make the detection environment at a vacuum. We will test a version with one vacuum cavity first, and later upgrade the detection system to a version with three vacuum cavities, as shown in Fig.~\ref{fig:vacuum}.

Additionally, we can combine several cosmic-ray scattering detectors as they can function simultaneously. This reduces the limitations of individual detectors and thus improves the accuracy of the detection so that we can better comprehend the reaction mechanisms of matter. Meanwhile, the use of multi-detector networking to form large-scale, wide-area cosmic-ray scattering detection arrays for accumulating data over long periods of time can lower the lower limit of dark matter detection.

\begin{figure}
    \centering
    \includegraphics[width=1.\columnwidth]{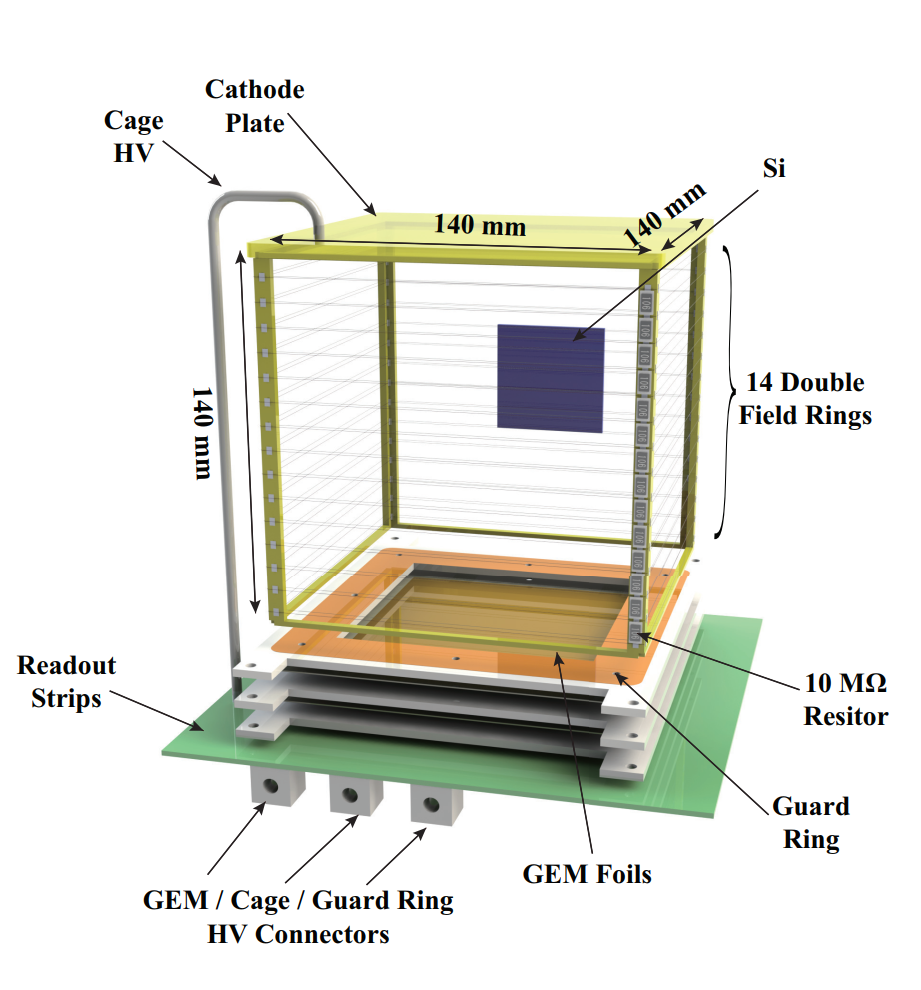}
    \caption{Schematic view of the PKU AT-TPC.}
    \label{fig:TPC}
\end{figure}

The Active-Target of Time Projection Chamber (AT-TPC) is a type of gas detector. It uses a working gas medium as the target material. The AT-TPC has advantages of recording the complete kinematic information and covering a large solid angle. Peking University has already built a small AT-TPC with a volume of $14\times14\times14~{\rm cm}^{3}$. It consists of an electronic field cage, double-layer GEM membrane for signal amplification and a 2-dimensional strip-readout structure. Its construction is shown as Fig.~\ref{fig:TPC}~\cite{Yang:NST}. In the future, we plan to increase its size to $25\times 25\times50~{\rm cm}^{3}$.

\begin{figure}
    \centering
    \includegraphics[width=1.\columnwidth]{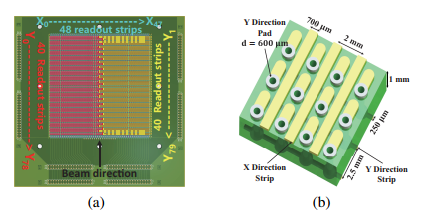}
    \caption{(a) Schematic view of readout PCB, and (b) amplified details of X-Y readout structure.}
    \label{fig:readout}
\end{figure}

A 2-dimensional strip-readout structure is shown as Fig.~\ref{fig:readout} \cite{Yang:NST}. To characterize the performance of the TPC, it is convenient to define a coordinate where the Z-axis is along the direction of the mean electron drift in the field cage and the X-Y axis lies in the readout plane. As shown in Fig.~\ref{fig:readout}(b), a total of 48 parallel X-strips with a width of 700~$\mu$m and 48 rows of pad plates were placed alternately. Each pad plate was extended to the back side of the board through a hole and collected on one of the Y-strips. In order to improve the positional resolution, each Y-strip was divided into two parts as shown in Fig.~\ref{fig:readout}(a), using a sawtooth cut away in the middle of the strip.

In our vision, an AT-TPC will be placed between RPCs. With 1-dimensional position readout plus a drift time dimension, we are able to get 2-dimensional tracks and energy loss. As high-speed muon’s ionizing capacity varies little before and after the collision, the variation of unit length energy loss mostly depends on the type of the other scattering particle---a charged particle will cause a high variation while an uncharged one (like neutrons and dark matter, but the muon’s energy is too low to knock out a neutron) will not. This method provides a further confirmation on whether a muon scattering event infers the dark matter’s existence.

\begin{figure}
    \centering
    \includegraphics[width=1.\columnwidth]{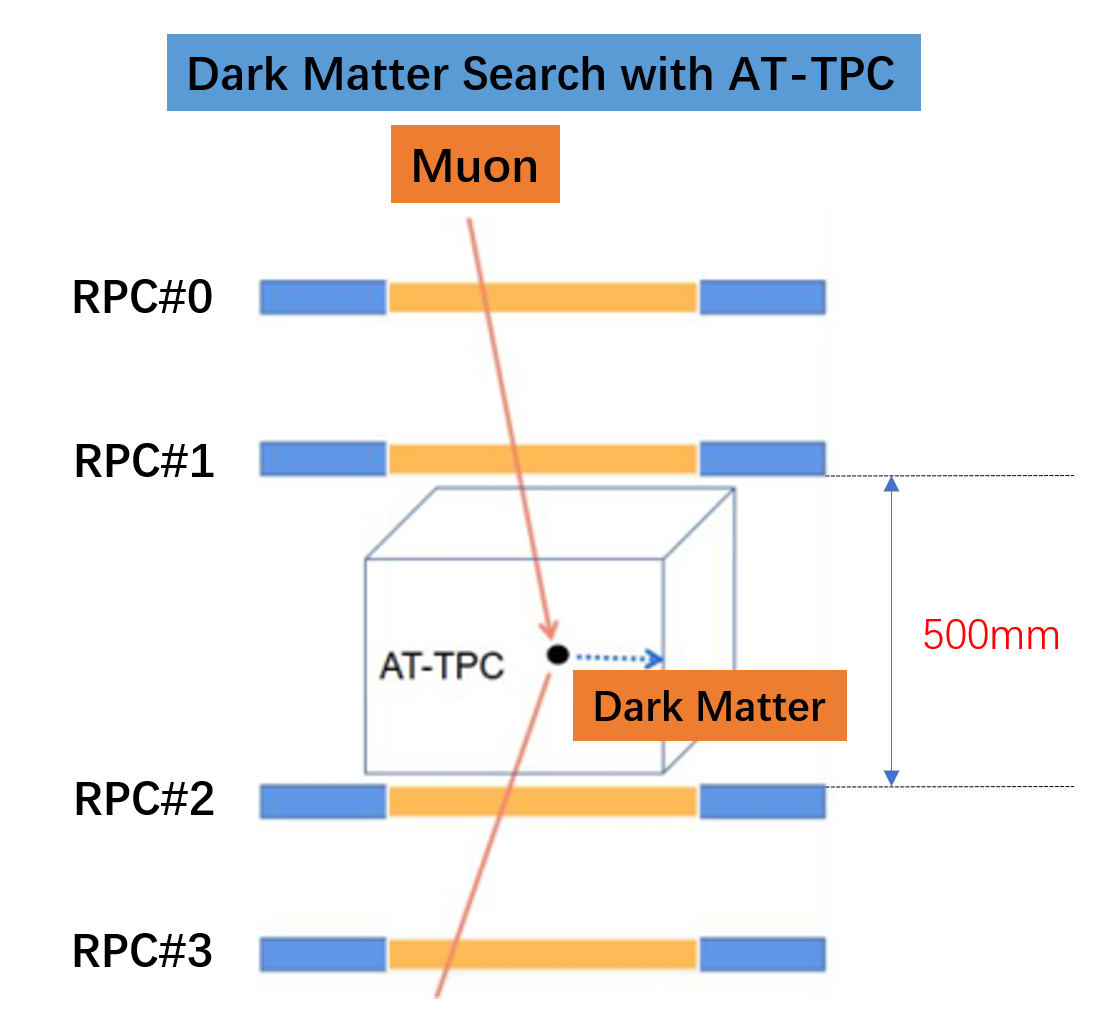}
    \caption{Schematic diagram of RPC testing system with AT-TPC.}
    \label{fig:RPCandTPC}
\end{figure}

Therefore, we will apply AT-TPC to get the track of particles for better precision, as shown in Fig.~\ref{fig:RPCandTPC}. Also, AT-TPC can work on some muon beamlines of relatively low energy, such as MuSIC in the RCNP~\cite{Cook:2017PRAB}. For the reason that some muons in relatively low energy ranging from a few MeV to tens of MeV cannot traverse RPC or will be scattered strongly, AT-TPC can be a suitable detector in this situation.

\section{Simulations Framework} 

The simulation of the GEM-based detector is conducted using \textsc{Geant4}~\cite{GEANT4:2002zbu}. The schematic design of the triple-GEM detector can be found in Ref.~\cite{gemd}. The drift cathode and readout PCB consist of 0.1~mm thick copper positioned on both sides of the GEM. The readout strip pitch is 210~$\mu$m, and the strip width is 150~$\mu$m. The GEM foil is composed of a 50~$\mu$m thick Kapton foil with 5~$\mu$m thick copper cladding on both sides. In the original GEM design, the foil is etched using photolithographic processes to form tiny bi-conical holes arranged in a regular hexagonal pattern. In our work, we approximate the avalanche amplification process that occurs when an electron passes through a hole with very high electric field strength. We achieve this by simulating the material effects of the tiny holes, reducing the density of the copper plate equivalently. The position resolution, including electron avalanche amplification, charge collection, readout strip pitch, $etc.$, is considered by a smearing according to the nominal position resolution. The thicknesses of the drift region, the three transfer regions, and the induction region are 4.8-2-2-2~mm. The upper and lower tracking systems consist of two stacked plates, between which is the $1\times 1\times 1~{\rm m}^{3}$ sensitive detection region in physics cases I and II below. The box is covered with a 100~$\mu$m thick Kapton shell. When tuned for dark matter searching mode, the box is filled with air or vacuum. In Tomography mode, the sample to be tested is placed inside.

\subsection{Dark Matter and muon scattering} 

The interaction of low-mass muon-philic DM interaction with free muons is illustrated in Figs.~\ref{fig:muonbox}, \ref{fig:mubeambox}, and \ref{fig:moutain}. In the presence of surrounding DM, these atmospheric or accelerator muons might experience spatial shifts, providing a distinct signal for detection. According to the standard halo model~\cite{Lewin:1995rx,Aalbers:2018mfc}, the velocity distribution of DM particles in the galactic rest frame follows a Maxwell-Boltzmann distribution with a cutoff at an escape velocity $v_{\mathrm{esc}}$, which is
\begin{equation}
    F_g\left(\mathbf{v}_{\mathbf{g}}\right)= \begin{cases}k\left(\frac{1}{\sqrt{\pi} v_0}\right)^3 \exp \left(-\left(v_g / v_0\right)^2\right) & v_g<v_{\mathrm{esc}} \\ 0 & v_g \geq v_{\mathrm{esc}}\end{cases},
\end{equation}
where $v_0=\sqrt{2K_B T/m}$ is the most probable or modal speed of the DM particles, and the normalization constant $k$ (1 if $v_{\mathrm{esc}}=\infty$) is
\begin{equation}
    k=\mathrm{erf}(v_{\mathrm{esc}}/v_0)-\frac{2}{\sqrt{\pi}}\frac{v_{\mathrm{esc}}}{v_0}\exp{-(v_{\mathrm{esc}}/v_0)^2}.
\end{equation}
As the Earth moves relative to the galactic rest frame with a velocity $v_E$, the velocity distribution of DM particles detected in an Earth-based detector is not isotropic. Defining the polar angle $\phi$ with its origin in the direction of the Earth’s motion, and the azimuth angle $\theta$ as the angle between the observed speed $v$ and the speed of the Earth $v_E$, the speed distribution is obtained by integrating over $\phi$ and $\theta$. The distribution is given by:
\begin{equation}
    f(v)=\int_0^{2 \pi} d \phi \int_0^\pi d \theta v^2 \sin (\theta) F_g\left(v_g\right).
\end{equation}
DM scattering experiments almost universally assume the standard halo model with parameters $v_0=220~\mathrm{km/s}$, $v_E=232~\mathrm{km/s}$, and $v_{\mathrm{esc}}=544~\mathrm{km/s}$ to ensure their results can be fairly compared with each other.
We also consider a simpler case by neglecting the velocity distribution of DM, taking it as a constant value $v=220~\mathrm{km/s}$.

In our work, we consider a model-independent elastic scattering between muons and DMs, following Newtonian mechanics. Conservation of energy and momentum can be used to show the muon recoil energy $E_{\mathrm{recoil}}$ obeys~\cite{Lewin:1995rx,Aalbers:2018mfc}
\begin{equation}
\begin{aligned}
    E_{\mathrm{recoil}}(v) &= \frac{{\rm M}_{\mathrm{DM}}v^2}{2} \frac{1-\cos\theta}{2}\frac{4{\rm M}_{\mathrm{DM}}{\rm M}_\mu}{({\rm M}_{\mathrm{DM}}+{\rm M}_\mu)^2} \\
    &=\frac{{\rm M}^{\prime}_{\mathrm{DM},\mu} v^2}{{\rm M}_\mu} (1-\cos\theta),
\end{aligned}
\end{equation}
where $\theta$ is the scattering angle in the center-of-mass frame and ${\rm M}^{\prime}_{\mathrm{DM},\mu}$ is the DM-muon reduced mass. The maximum recoil energy is achieved at a head-on collision ($\theta=\pi$):
\begin{equation}
    E_{\mathrm{recoil}}^{\mathrm{max}} = \frac{2{\rm M}^{\prime}_{\mathrm{DM},\mu} v^2}{{\rm M}_\mu}.
\end{equation}

The incident cosmic-ray muons are simulated by CRY package~\cite{CRY}. Since the mean energy of cosmic-ray muons at sea level is about 3-4~GeV, making them relativistic, DM is quasi-frozen from the perspective of fast incoming muons. This differs from other DM search experiments, such as XENON1T and PandaX, where nuclei are quasi-static target and DMs are assumed to be incoming particles. Notice for high speed muons, it is appropriate to treat DM as frozen in the detector volume, and the estimated muon-DM scattering rate per second could be, for the case of cosmic-ray muons:
\begin{equation}
dN/dt=\rho V/{\rm M}_{\rm DM} \times \sigma_{\mu,\mathrm{DM}} \times F_{\mu}. \nonumber
\end{equation} 
Here, $F_{\mu}$ is the muon flux $\sim$ 1/60~s$^{-1}$cm$^{-2}$ at sea level. The local density of DM is on the order of $\rho\sim$~0.3~GeV/$\rm{cm}^3$, and the number of DM in a box with volume $V$ is $N_{\mathrm{DM}} = \frac{\rho V}{{\rm M}_{\mathrm{DM}}}$. For the DM mass ${\rm M}_{\mathrm{DM}}\sim1$ (0.1, 10)~GeV, and the detector box volume $V\sim 1\,{\rm m}^3$, the number of DMs is about $3\times10^{5(6,4)}$. The sensitivity of DM-muon scattering cross section for 1 year run will be around
\begin{equation}
\sigma_{\mu,\mathrm{DM}} \sim 10^{-12(-13,-11)}~{\rm cm}^2.  \label{eq:muonbox}
\end{equation} 

On the other hand, in the case of muon beams, we have:
\begin{equation}
dN/dt=N_{\mu}\times \sigma_{\mu,{\mathrm{DM}}} \times L\times \rho /{\rm {\rm M}_{\rm DM}}. \nonumber
\end{equation} 
Taking the example of ${\rm M}_{\mathrm{DM}}=0.03$\,GeV, $L=1$\,m, and $N_\mu\sim 10^6$/s ($e.g.$, China CSNS Melody design), we get the following estimation:
\begin{equation}
N=10^{13}\times \sigma_{\mu,\mathrm{DM}} \times 100/{\rm cm}^2, \label{eq:muonbeambox}
\end{equation} 
Thus the sensitivity on DM-muon scattering cross section for 1 year run will be around 
\begin{equation}
\sigma_{\mu,\mathrm{DM}} \sim 10^{-15}~{\rm cm}^2. \nonumber
\end{equation} 

Notice, however, detector acceptance and efficiency needs to be applied, which can change above estimated results and will be further explained in the sections below.

Moreover, our methods can have advantages over `exotic' DMs~\cite{McKeen:2023ztq} which can be well slowed down through scattering with matter in the atmosphere or the Earth before reaching the detector target. In such a scenario, dark matter number density can be as large as $10^{15}$ cm$^{-3}$, and sensitivity on dark matter and muon scattering cross section can reach near microbarn level.

\section{Physics Case I:  Precise measurements of cosmic-ray muons and Muon Tomography}
\label{caseI}

\begin{figure}
    \centering
    \includegraphics[width=1.\columnwidth]{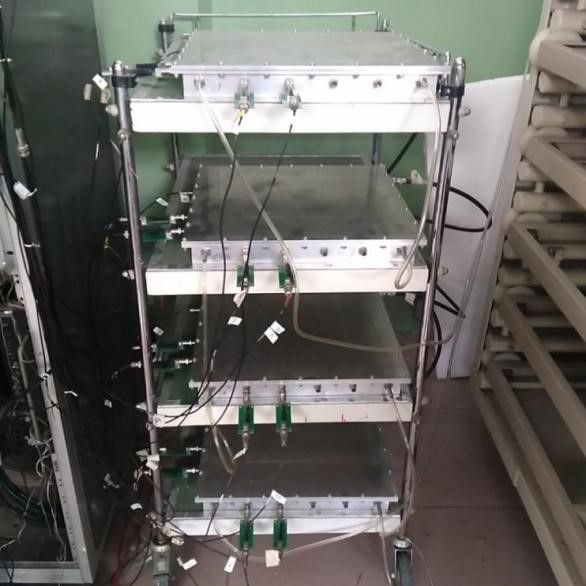}
    \caption{Prototype of muon imaging system based on delay line readout and glass RPC developed by Peking University.}
    \label{fig:prototype}
\end{figure}

\begin{figure}
    \centering
    \includegraphics[width=1.\columnwidth]{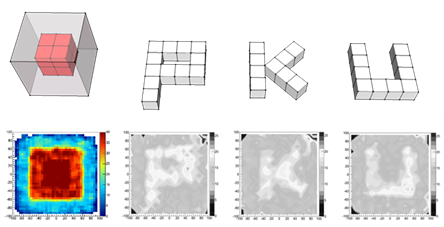}
    \caption{Imaging results of a $6\times6\times6~{\rm cm}^{3}$ square lead block wrapped in a $12\times12\times12~{\rm cm}^{3}$ iron shell and the letters PKU composed of $3\times3\times3~{\rm cm}^{3}$ iron blocks by muon imaging prototype of Peking University. The color represents the average scattering angle of the corresponding region.}
    \label{fig:imaging result}
\end{figure}

The first differential and integral spectra of muons moving in the near-vertical direction can be measured at Beijing using the devices as mentioned above. As the first step, we put 4 RPCs in our muon tomography systems, as illustrated in Fig.~\ref{fig:prototype}. When cosmic-ray muons traverse materials, they undergo multiple deflections. The test results, depicted in Fig.~\ref{fig:imaging result}, clearly show the outline of the lead block in the iron shell, indicating differences in muon scattering angles as they pass through various materials.

The idea of muon tomography is based on the fact that cosmic-ray muons will undergo Coulomb scattering when they pass through materials. Therefore, by measuring the muon scattering angular distribution, the distribution information of substances or materials in the detection area can be reconstructed. In 2004, Schultz \textit{et al.} first conducted such experiments and proposed Point of Closest 
Approach (PoCA) algorithm~\cite{Schultz:2004kx}, in which multiple consecutive scatterings of muons are approximated as one scattering, and the scattering point (called ``PoCA'' point) is located at the midpoint of the common perpendicular line between the incident and outgoing straight lines. Subsequently, in 2006-2007, Schultz \textit{et al.} proposed Maximum Likelihood Scattering and Displacement (MLSD) algorithm~\cite{MLSD1,MLSD2}, a new iterative imaging algorithm based on the maximum likelihood principle in statistics. This algorithm is proposed to take into account the muons scattering angle and then introduce the influence of the lateral displacement during the muons penetration process into the material. However, these algorithms have certain technical limitations. To date, new muon tomography imaging algorithms continue to be proposed, for example, the Ratio algorithm based on the PoCA algorithm~\cite{Cheng-Ming:2019rjf}.

\section{Physics Case II:  Dark matter searches in a box}
\label{caseII}

For muon-philic DM, one can start with a relatively straightforward searches using those free cosmic-ray muons, as depicted in Fig.~\ref{fig:muonbox}. A quick estimation of the projected sensitive on DM muon scattering cross section is shown in Eq.~\ref{eq:muonbox}, which needs further refinement to incorporate detailed simulation effects, as explained blow.

\begin{figure}
    \centering
    \includegraphics[width=1.\columnwidth]{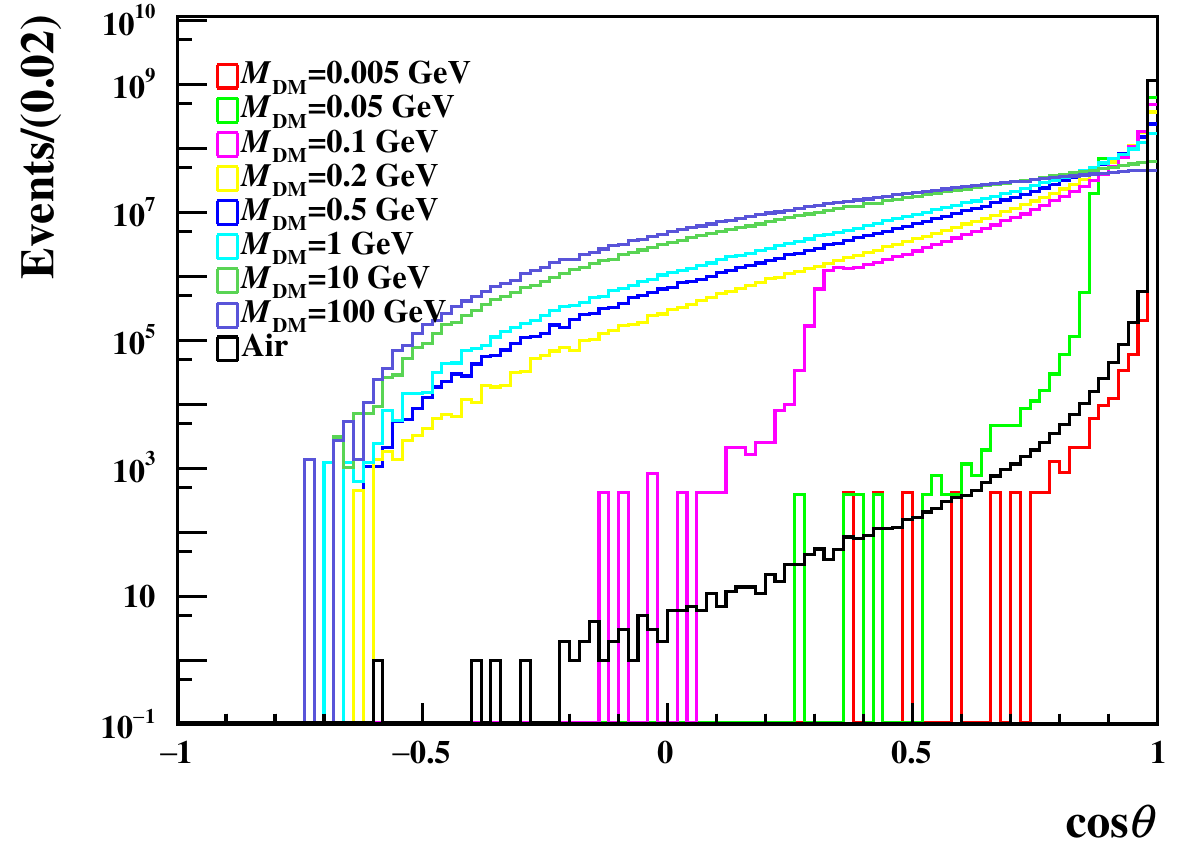}
    \caption{The $\cos\theta$ distributions in signal and background samples. The black histogram is the background, while the other colored histograms denote signal under different DM mass assumptions, scaled to the same event number. In the large scattering region, $i.e.$, $\cos\theta<0$, signal distributions with a large DM mass and background distributions manifest a noticeable difference.}
    \label{fig:costheta_box}
\end{figure}

We simulate 5 billion incident muon events to study the backgrounds in one-year exposure, and also simulate 10 million signal events under different DM assumptions to estimate the signal efficiency. Due to the low cross section of DM-muon interaction, we assume that there is only one interaction in the box. We require a muon hit signal in each GEM detector. The MC truth hit position, smeared by the GEM spatial resolution is taken as the reconstructed hit position. The two hits in the upper (lower) GEM detector determine the direction of incoming (outgoing) muons. The angle between the directions of incoming and outgoing muon, $\theta$, is the discriminant variable. The $\cos\theta$ distributions in signal and background samples are shown in Fig.~\ref{fig:costheta_box}. The signal distributions are scaled to the event numbers in the background. We list some interesting results as below.
\begin{itemize}
    \item The $\cos\theta$ distribution in air has no obvious difference between that in a vacuum. Therefore, the scattering between muons and air has no significant effect under the current GEM spatial resolution. Considering cost and technical difficulty, vacuuming of the boxes is not necessary in Phase I of the project.
    \item The $\cos\theta$ distributions in Maxwell-Bolzmann velocity distribution and a constant velocity distribution are similar. Therefore, our signal distribution and detection is not sensitive to the DM velocity model.
    \item As the DM mass increases, a larger fraction occupies the region of large scattering angles, resulting a more pronounced discrepancy between the signal and background.
    \item For the signal event with ${\rm M}_{\mathrm{DM}}<100$~MeV, an apparent truncation is observed, attributed to kinematics. This truncation occurs only when the DM mass is lower than the muon mass.
\end{itemize}
To further suppress backgrounds, we require that the reconstructed PoCA point should be within the box, which is reasonable and supported by Punzi Figure-of-Merit, $\mathrm{FoM}=\varepsilon/(3/2+\sqrt{B})$~\cite{Punzi:2003bu}. Here, $\varepsilon$ is the detection efficiency, and, and $B$ is the event number of background events. The FoM value is improved by 18\%. The resulting detection efficiency under different DM mass assumptions and the background event numbers are listed in Table.~\ref{tab:bkgnum_sigeff}.

\begin{table}[htbp]
    \centering
    \begin{tabular}{c|cc}
        \hline\hline
         \text{Background} & \multicolumn{2}{c}{\text{Event Number }($\times10^{9}$)} \\ \hline
         \text{Air} & \multicolumn{2}{c}{1.15} \\ \hline
         \text{Vacuum} & \multicolumn{2}{c}{1.14} \\
        \hline\hline
          \text{DM mass (GeV)} & \text{Constant (\%)} & \text{Maxwell-Bolzmann (\%)} \\ \hline
          0.005 & $27.10\pm0.01$ & $27.11\pm0.01$ \\ \hline
          0.05 & $29.56\pm0.01$ & $29.55\pm0.01$ \\ \hline
          0.1 & $27.66\pm0.01$ & $27.64\pm0.01$ \\ \hline
          0.2 & $25.01\pm0.01$ & $24.99\pm0.01$ \\ \hline
          0.5 & $21.47\pm0.01$ & $21.46\pm0.01$ \\ \hline
          1 & $18.67\pm0.01$ & $18.66\pm0.01$ \\ \hline
          10 & $11.10\pm0.01$ & $11.10\pm0.01$ \\ \hline
          100 & $8.44\pm0.01$ & $8.43\pm0.01$ \\
        \hline\hline
    \end{tabular}
    \caption{Background event numbers corresponding to the integrated luminosity of one-year exposure with the box filled with air and vacuum, along with the signal detection efficiency under different assumptions of DM velocity distribution and mass.}
    \label{tab:bkgnum_sigeff}
\end{table}

\begin{figure}
    \centering
    \includegraphics[width=1.\columnwidth]{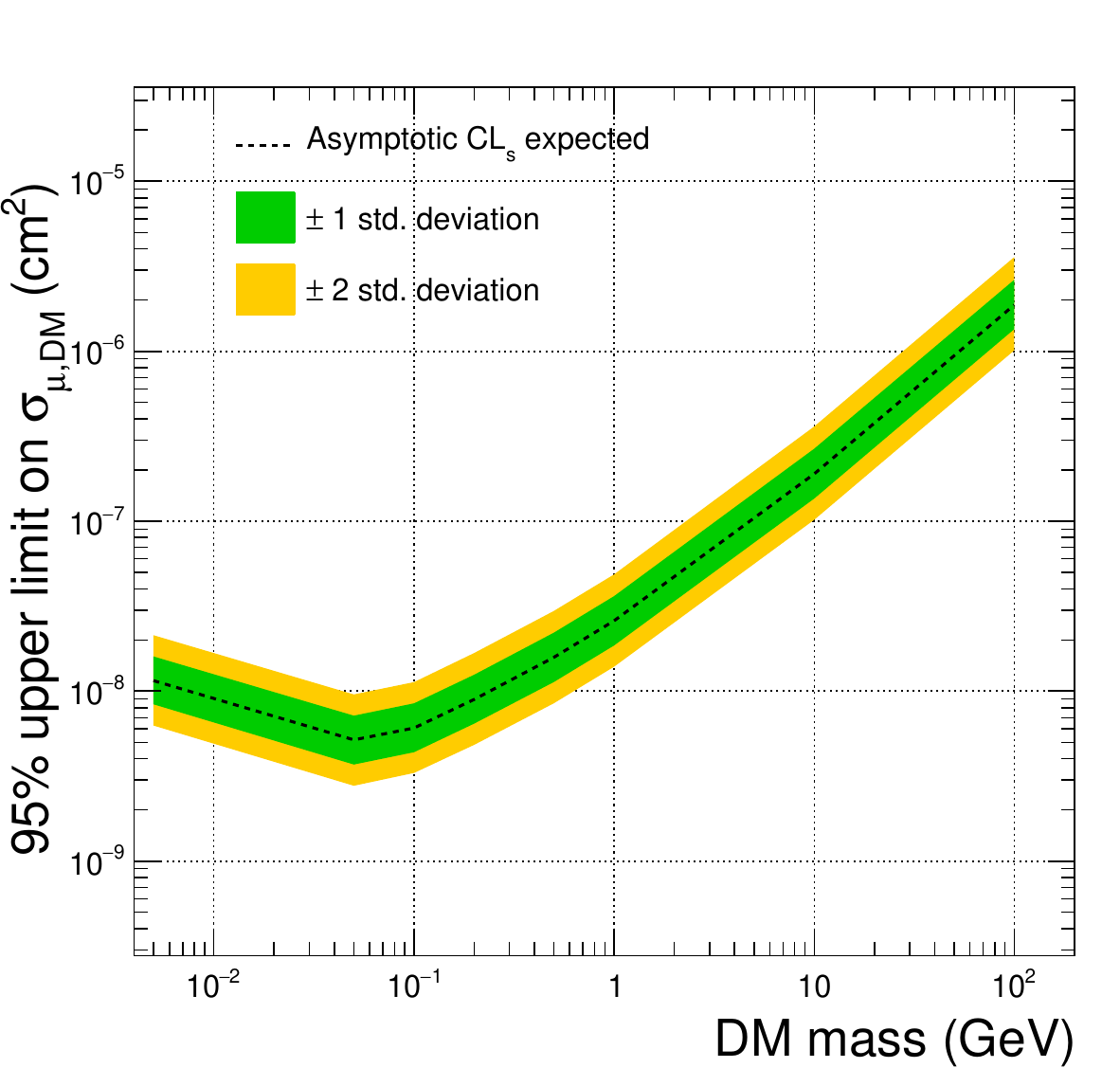}
    \caption{Expected 95\% CL upper limits on the DM-muon interaction cross section versus DM mass. The green and yellow bands denote $1\sigma$ and $2\sigma$ region, respectively.}
    \label{fig:UL_box}
\end{figure}

We search for excesses in the $\cos\theta$  distribution at the lower range by performing binned maximum likelihood fits using Higgscombine~\cite{Higgscombine}. As observed yields are unavailable for this R\&D study, ``Asimov'' data is used. We set the upper limit (UL) using the CLs technique~\cite{Junk:1999kv,Read:2002hq}. Assuming the measurements performed are limited by the availability of data statistics, have negligible experimental systematic, and have excellent MC statistics, the background uncertainty is taken as the Poisson counting uncertainty for the expected background yield in each bin. The upper limit of the DM-muon interaction cross section at 95\% confidence level (CL) corresponding to an integrated luminosity of one year are shown in Fig.~\ref{fig:UL_box}. Taking into account the detection efficiency, the upper limits on cross section for sub-GeV DM and muon interaction are set to $10^{-7}\sim10^{-9}\,\mathrm{cm}^{2}$.

Our methods can have advantages over `exotic' DMs~\cite{McKeen:2023ztq} which can be well slowed down through scattering with matter in the atmosphere or the Earth before reaching the detector target. In such a scenario, dark matter number density can be as large as $10^{15}$ cm$^{-3}$, and sensitivity on dark matter and muon scattering cross section can reach as low as $10^{-22}\sim 10^{-24}\,\mathrm{cm}^{2}$.

\section{Physics Case III: Dark matter searches using muon beam}
\label{caseIII}

As estimated with Eq.~\ref{eq:muonbeambox} and shown in Fig.~\ref{fig:mubeambox}, owing to a much larger muon intensity and focused beam, we anticipate that the detector can be made more compact, leading to a further improvement in sensitivity for DM searches.

\begin{figure}
    \centering
    \includegraphics[width=1.\columnwidth]{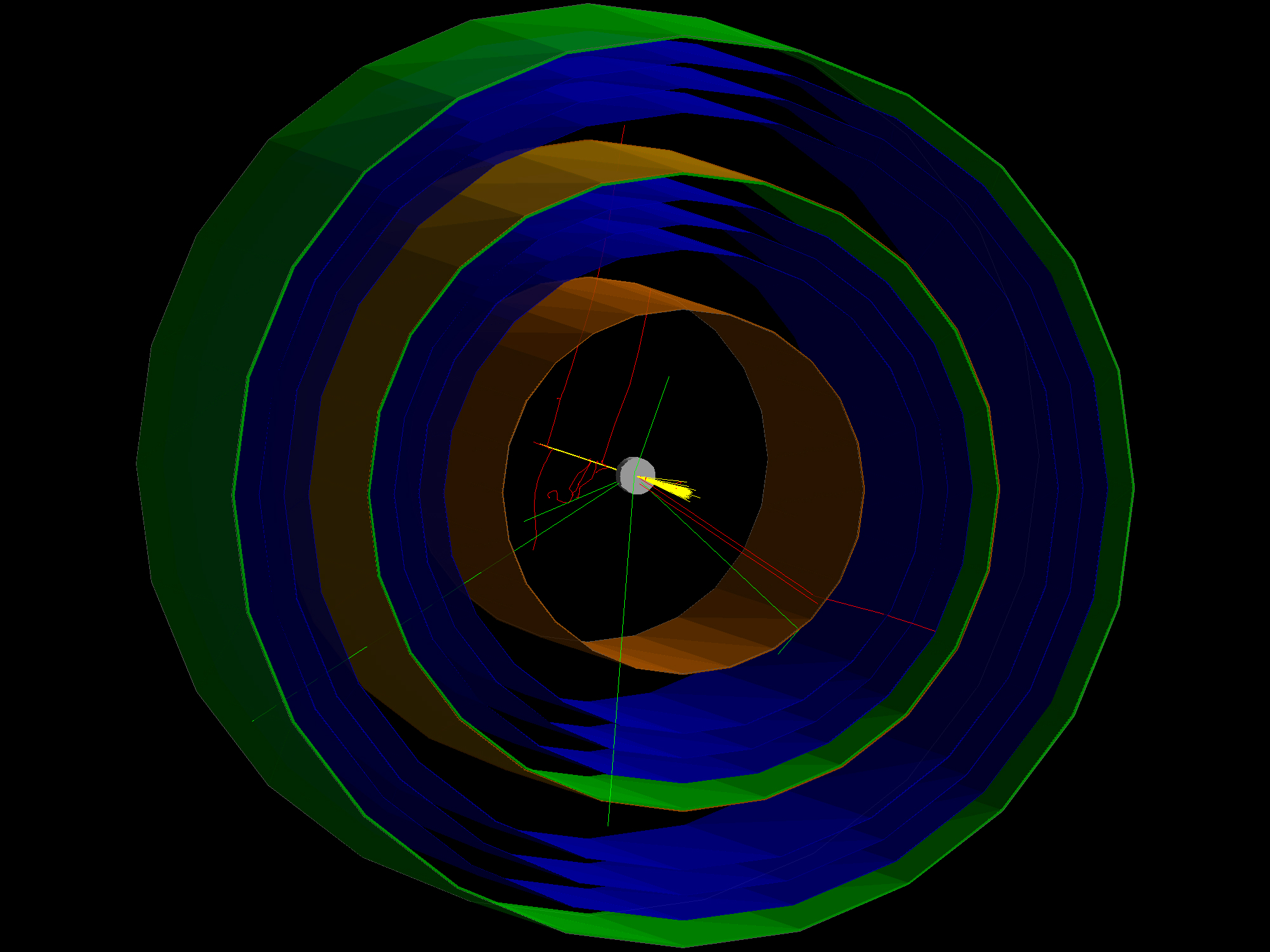}
    \caption{Simulating 1~GeV muon beam hit lead plate passing through GEM detector. Orange surfaces are drift cathodes. The blue surfaces are GEM foils. The green surfaces are PCBs. The yellow lines are muons tracks. The red curves are electron tracks. The green lines are photon.}
    \label{fig:mubeamG4vis}
\end{figure}

To validate the correctness of estimation, we perform a MC simulation study on DM searches using a muon beam via {\sc Geant4}~\cite{GEANT4:2002zbu}. As shown in Fig.~\ref{fig:mubeamG4vis}, to suit detection environment of the beam experiment, we adopt cylindrical GEM (CGEM) detector structure. This innovative technique has been used in the upgrade of the BESIII inner tracker system~\cite{Bortone:2022ktq}. According to the MELODY design~\cite{Tang:2010zz,Bao:2023nup}, the diameter of the beam spot, $\phi$, ranges from $10\,\mathrm{mm}$ to $30\,\mathrm{mm}$. In our study, we choose $\phi=10\,\mathrm{mm}$. To reduce beam background related to the size of the beam spot, the inner diameter of our CGEM detector is designed to be $50\,\mathrm{mm}$, which is 5 times the beam spot. The two layers of GEM detectors are stacked together, and the structure of each GEM is the same as described above. The length of the GEM detector is designed to be 1~m. For muon beam, we consider four different cases with low muon momentum $P_\mu=10\,\mathrm{MeV}$, $100\,\mathrm{MeV}$ and high muon kinetic energy $E^\mu_{\mathrm{kin}}=1\,\mathrm{GeV}$ and $10\,\mathrm{GeV}$. The momentum direction of the muon is along the $z$ axis, which is the symmetry axis of the detector, and the profile of beam in the $xy$ plane follows a Gaussian distribution.

\begin{figure*}[htbp]
    \centering
    \includegraphics[width=0.65\columnwidth]{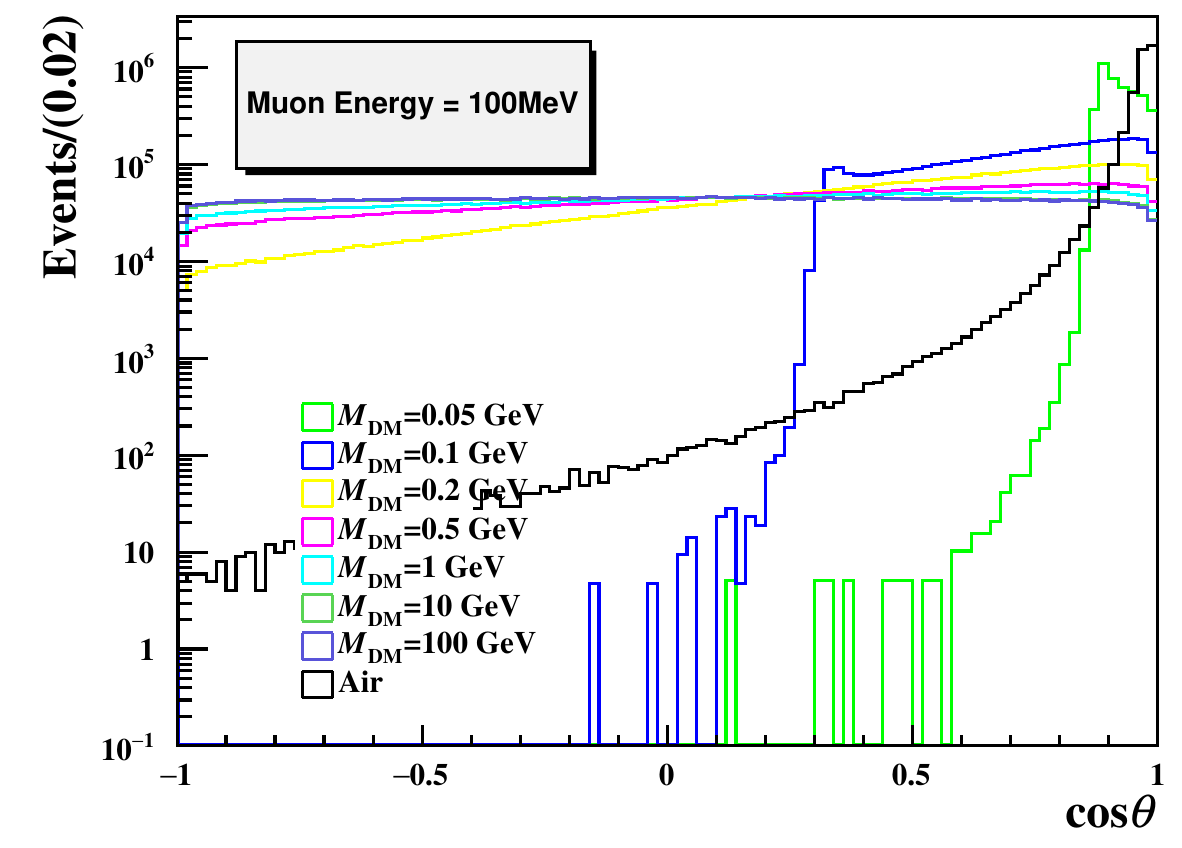}
    \includegraphics[width=0.65\columnwidth]{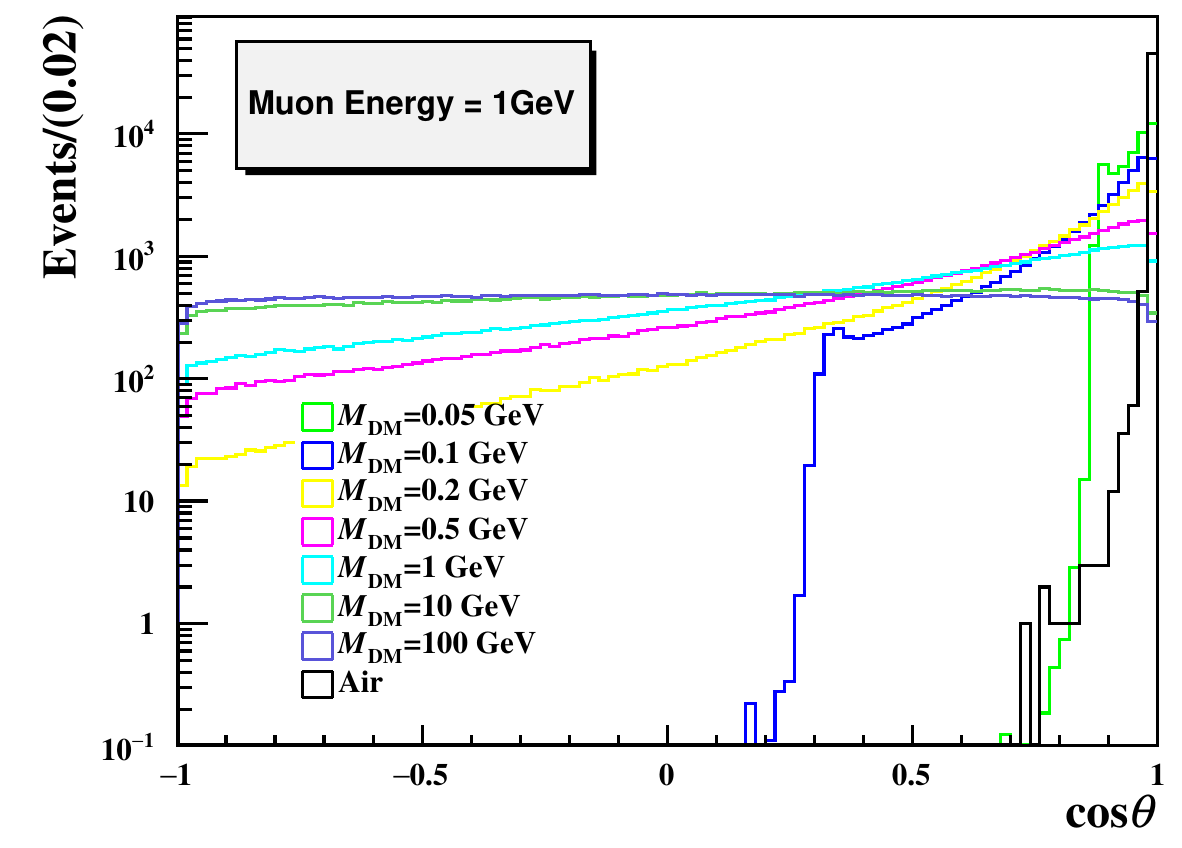}
    \includegraphics[width=0.65\columnwidth]{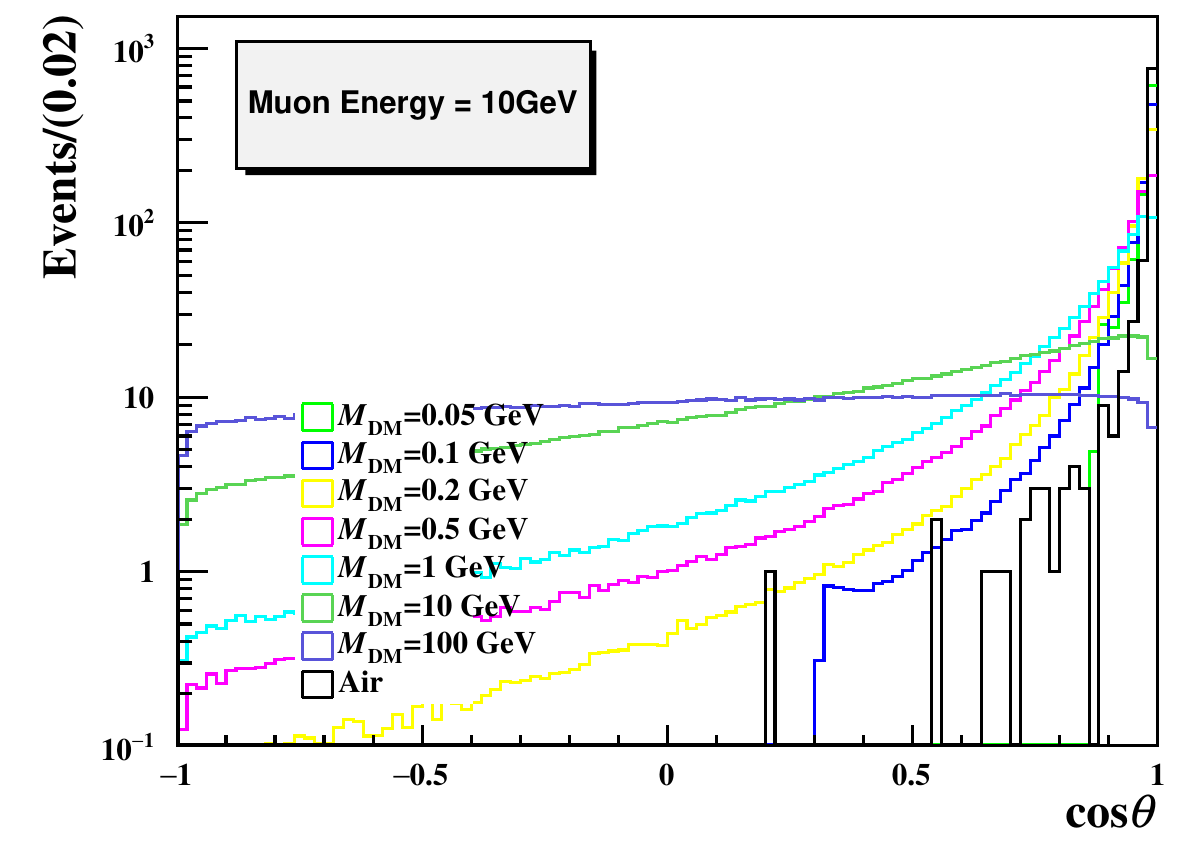}
    \caption{The $\cos\theta$ distributions with muon energies of 100~MeV, 1~GeV and 10~GeV in signal and background samples. The black histogram is the background, while the colored histograms denote the signal under different DM mass assumptions scaled to the same event number. In the large scattering region, $i.e.$, $\cos\theta<0$, the signal with a large DM mass and background distributions manifest a remarkable difference.}
    \label{fig:costheta_beam}
\end{figure*}

For the background case, we assume that the incoming muons scatter with air in the beam pipe, resulting in changes in their momenta. If the scattering angle is large enough, muons may hit the detector. The two hit positions in the two GEM detectors can reconstruct the flying direction of the outgoing muon. For signal case, we assume that the incoming muons experience one elastic scattering with DM particles in the detection region. The signal model is described above, and here we only consider the case of a DM Maxwell-Bolzmann velocity distribution. Figure~\ref{fig:costheta_beam} shows the $\cos\theta$ distributions of signal and background samples. Notably, since $P_\mu=10\,\mathrm{MeV}$ is too low, muon will lose all its energy scattering with air, resulting in no events surviving in the background simulation. This supports the idea that the beam pipe should be in a vacuum. However, in other cases where muon momentum is large enough, muons will hit the GEM detector. The resulting detection efficiency under different
DM mass assumption and different muon beam energies are listed in Table~\ref{tab:sigeff_beam}.

\begin{figure*}
    \centering
    \includegraphics[width=0.65\columnwidth]{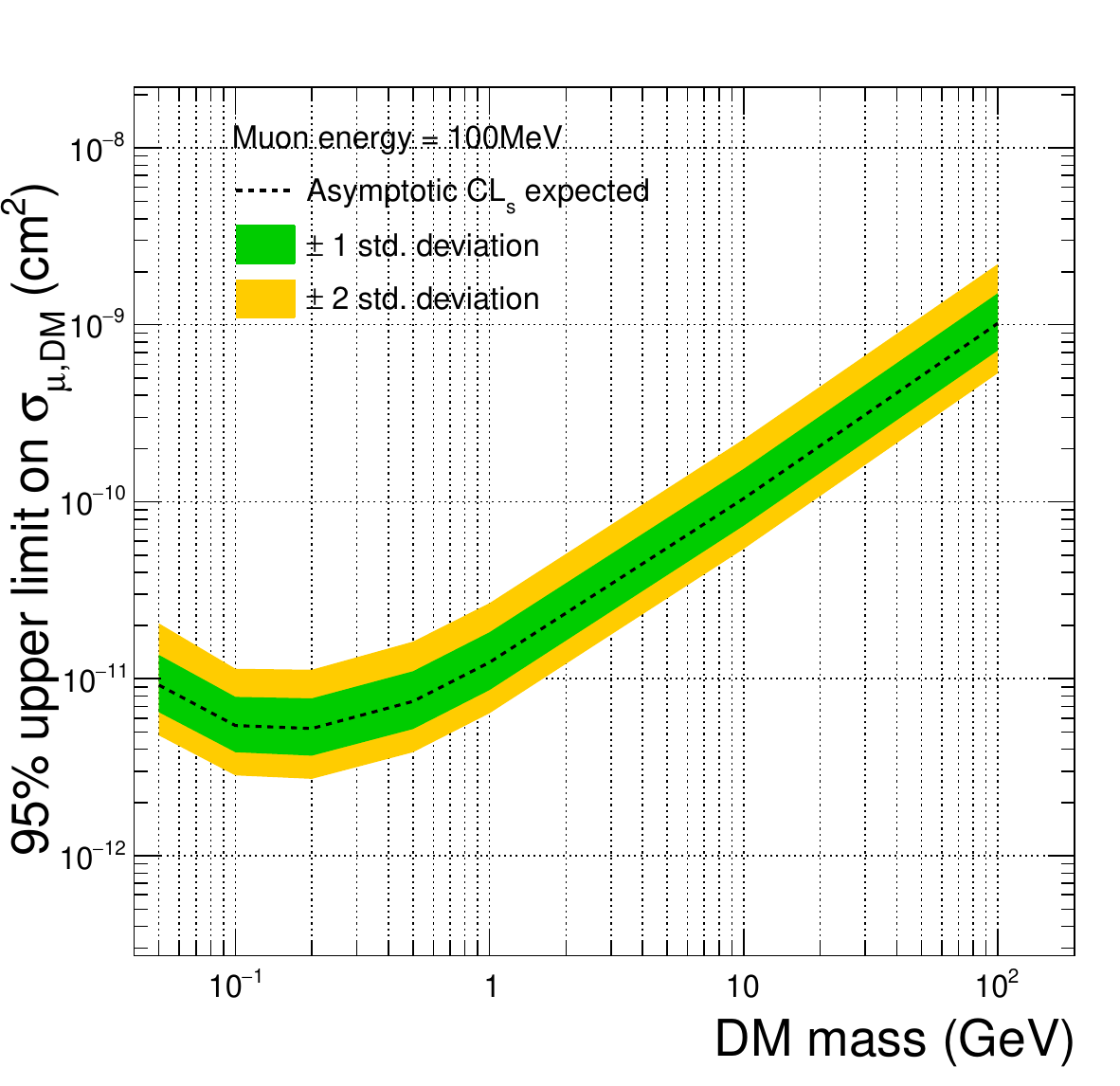}
    \includegraphics[width=0.65\columnwidth]{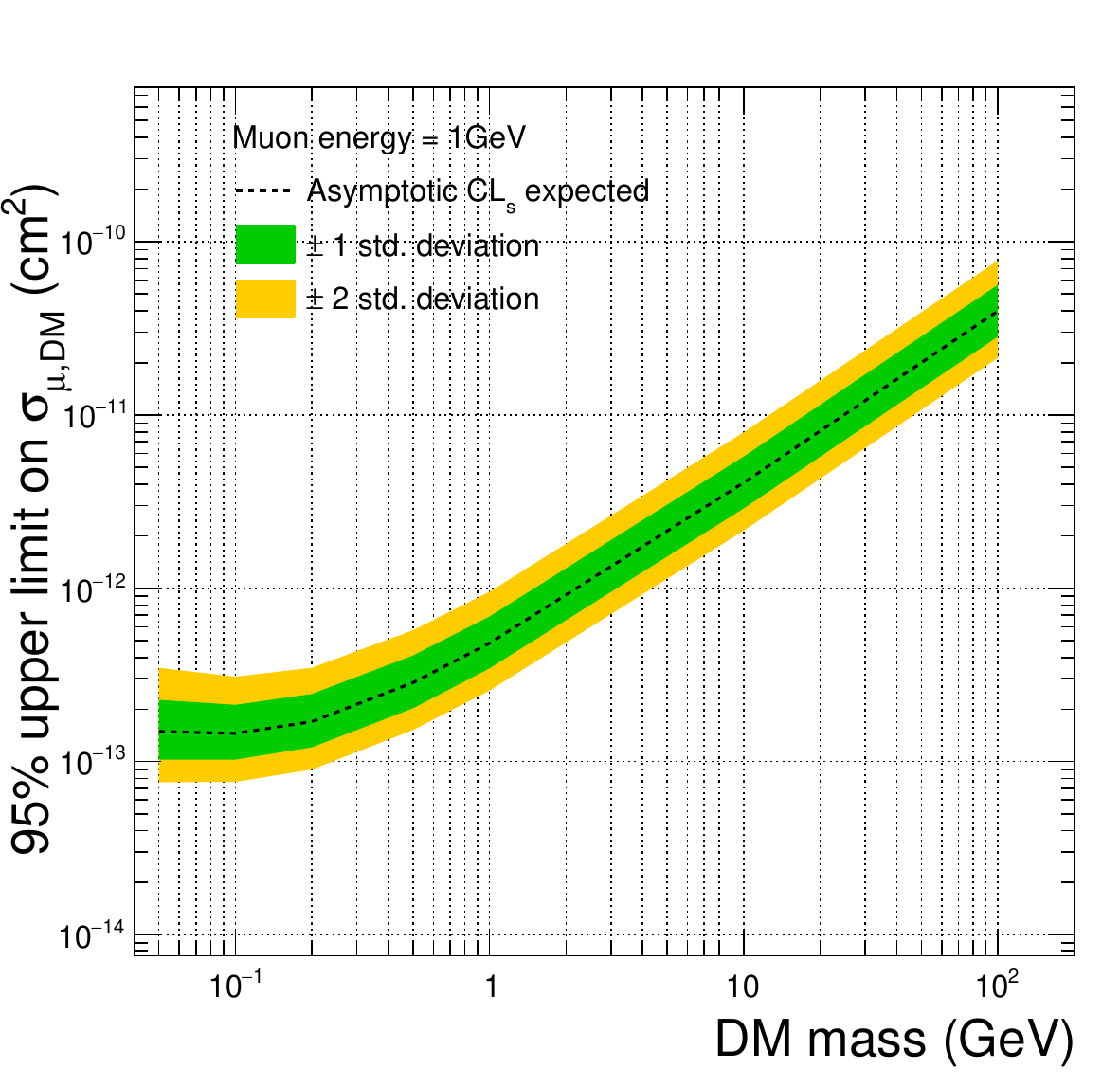}
    \includegraphics[width=0.65\columnwidth]{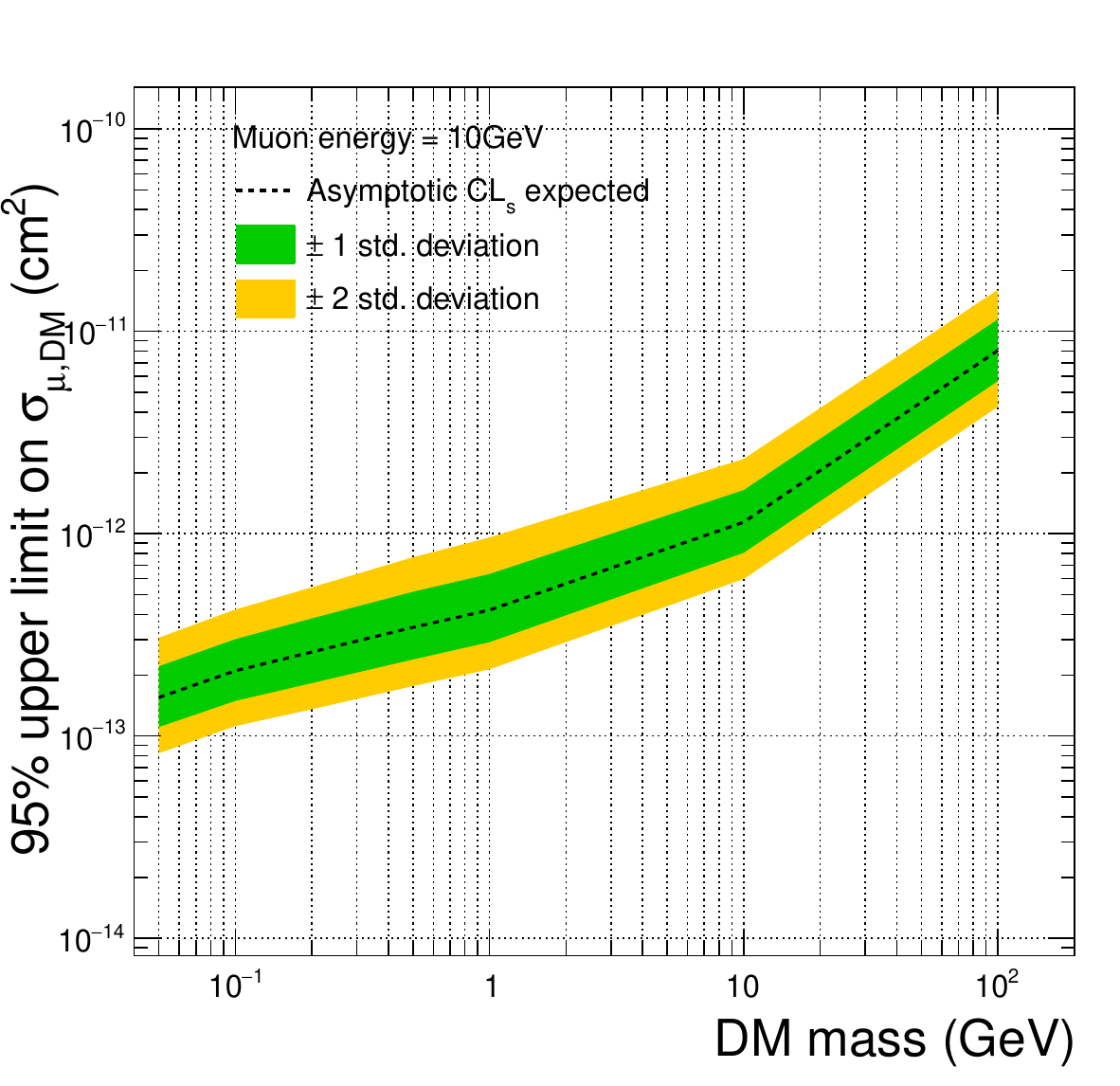}
    \caption{Expected 95\% CL upper limits on the DM-muon interaction cross section versus DM mass with muon beam energy 100~MeV, 1~GeV and 10~GeV. The green and yellow bands denote $1\sigma$ and $2\sigma$ region, respectively.}
    \label{fig:UL_beam}
\end{figure*}

We search for excesses in the $\cos\theta$ distributions by performing binned maximum likelihood fits using Higgscombine~\cite{Higgscombine}. Since observed yields are unavailable, ``Asimov'' data is used. We set the upper limit (UL) using the CLs technique~\cite{Junk:1999kv,Read:2002hq}. Assuming the measurements performed are limited by the availability of data statistics, have negligible experimental systematic, and have excellent MC statistics, the background uncertainty is taken as the Poisson counting uncertainty for expected background yield in each bin. The upper limit of the DM-muon interaction cross section at 95\% confidence level (CL) corresponding to an integrated luminosity of one year are shown in Fig.~\ref{fig:UL_beam}. Here, we consider a muon intensity is $I=10^6\,\mathrm{s}^{-1}$, consistent with the CSNS Melody design. Taking account detection efficiency, the upper limits on the cross section for sub-GeV DM and muon interaction are set to $10^{-9}\sim10^{-12}\,\mathrm{cm}^{2}$ when muon momentum is 100~MeV, $10^{-11}\sim10^{-13}\,\mathrm{cm}^{2}$ when muon energy is 1~GeV, and $10^{-12}\sim10^{-13}\,\mathrm{cm}^{2}$ when muon energy is 10~GeV.

Again, our methods can have advantages over `exotic' DMs~\cite{McKeen:2023ztq} which can be well slowed down through scattering with matter in the atmosphere or the Earth before reaching the detector target. In such a scenario, dark matter number density can be as large as $10^{15}$ cm$^{-3}$, and sensitivity on dark matter and muon scattering cross section can reach near microbarn level.

\begin{table}[htbp]
    \centering
    \begin{tabular}{c|ccc}
        \hline\hline
          \text{${\rm M}_{\rm DM}$ \textbackslash $E^\mu_{\mathrm{kin}}$}  & 100~MeV~(\%) & 1~GeV~(\%) & 10~GeV~(\%) \\ \hline
          %0.005 & $4.99\pm0.02$ & $1.79\pm0.01$ & $0.60\pm0.01$ \\ \hline
          0.05 GeV & $84.29\pm0.04$ & $74.85\pm0.04$ & $45.93\pm0.05$ \\ \hline
          0.1 GeV & $91.74\pm0.03$ & $83.07\pm0.04$ & $58.17\pm0.05$ \\ \hline
          0.2 GeV & $94.35\pm0.02$ & $88.16\pm0.03$ & $68.37\pm0.05$ \\ \hline
          0.5 GeV & $95.17\pm0.02$ & $92.16\pm0.03$ & $78.91\pm0.04$ \\ \hline
          1 GeV & $95.34\pm0.02$ & $93.88\pm0.02$ & $84.68\pm0.04$ \\ \hline
          10 GeV & $95.35\pm0.02$ & $95.36\pm0.02$ & $94.06\pm0.02$ \\ \hline
          100 GeV & $95.43\pm0.02$ & $95.37\pm0.02$ & $95.37\pm0.02$ \\
        \hline\hline
    \end{tabular}
    \caption{Signal detection efficiency under different assumptions of DM mass and muon beam energies.}
    \label{tab:sigeff_beam}
\end{table}

\section{Physics Case IV:  Dark matter searches between mountain and sea level}
\label{caseIV}

Directional distributions of cosmic-ray muons can be precisely measured, either at mountain or sea level, as shown in Fig.~\ref{fig:moutain}, and their differences may reveal possible information of DM distributed near the earth. 

\begin{figure}
    \centering
    \includegraphics[width=1.\columnwidth]{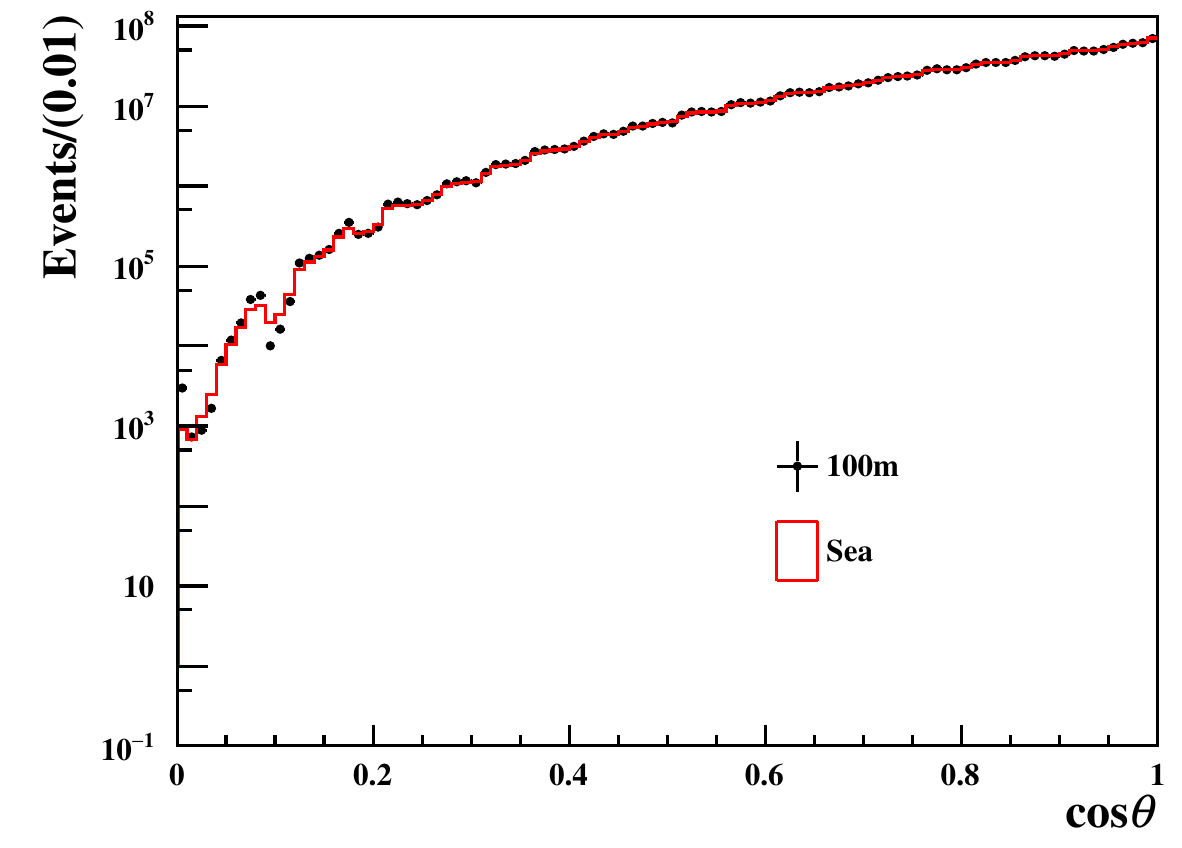}
    \caption{The $\cos\theta$ distributions at the mountain and sea levels in the background sample.}
    \label{fig:costheta_mountain_sea}
\end{figure}

In the {\sc Geant4} framework, the dominant physics process of muon-air interaction is ionization, in which the energy loss of muon is described by the Bethe-Bloch formula~\cite{pdg2023}. When traveling in the atmosphere, muons scatter with electrons of air molecule, and their momentum directions change accordingly. Due to limitations in computing resources, we simulate $1.5\times10^{9}$ events, which is approximately $30\%$ of the statistics for one year of data from a 1~m$^{2}$ detector. We place two layers of Gem detectors, both at the sea level and in a mountain with a latitude of 100~m, to measure the flying direction of cosmic-ray muons. The $\cos\theta$ distributions of muons at the mountain and sea levels are shown in Fig.~\ref{fig:costheta_mountain_sea}.

\begin{figure}
    \centering
    \includegraphics[width=1.\columnwidth]{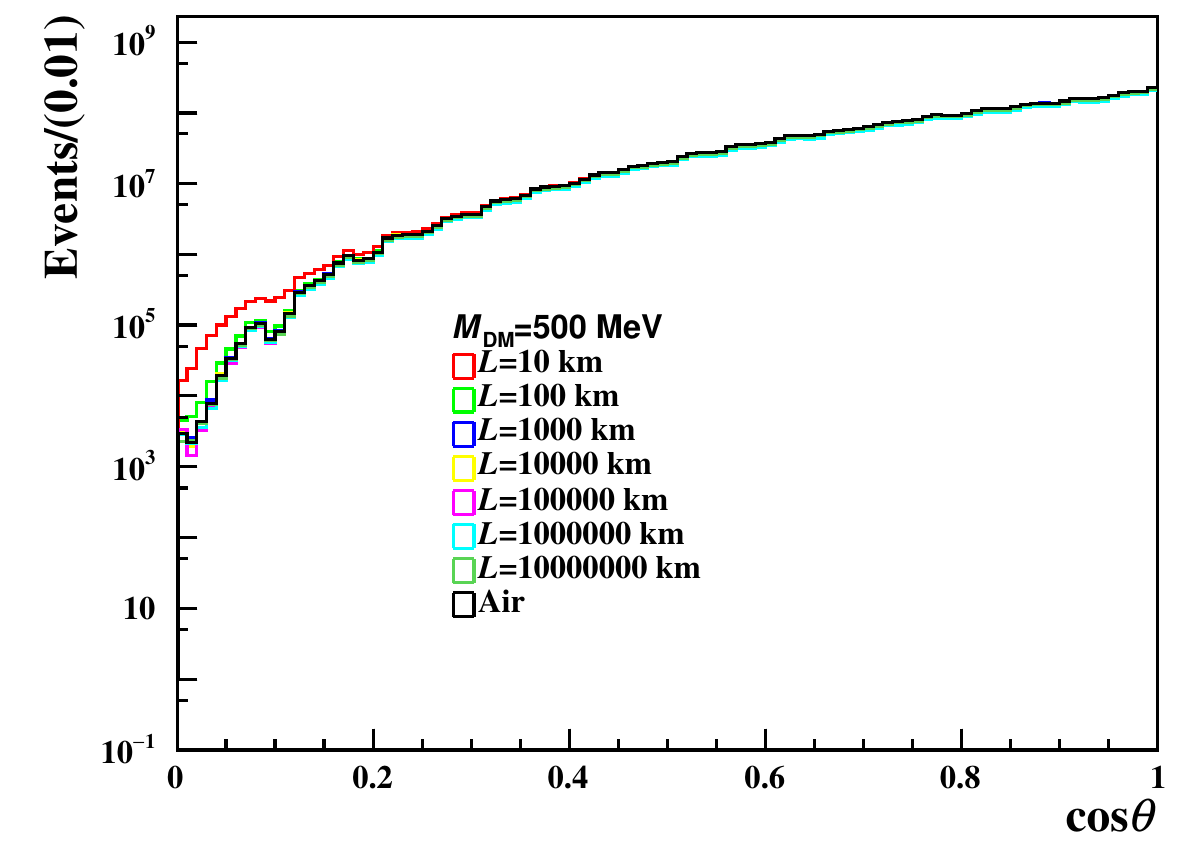}
    \caption{The $\cos\theta$ distributions at the sea level in different signal samples. The DM mass is 500~MeV, and the mean free path ranges from $L=10^1$~km to $L=10^7$~km. The histograms have been scaled to one year statistics.}
    \label{fig:costheta_signal_sea}
\end{figure}

Assuming DM clouds surround the earth, there is a probability that cosmic-ray muons scatter with DM particles when traveling through the air. To simulate signal MC samples, we extend the native {\sc Geant4} physics list by introducing the muon-DM elastic scattering process. The cross section of muon-DM elastic scattering process can be set manually by mean free path $L$. The relation between muon-DM cross section and mean free path is given by $L=1/(n_{\mathrm{DM}}\times\sigma_{\mu,\mathrm{DM}})$, where $n_{\mathrm{DM}}=\rho_{\mathrm{DM}}/{\rm M}_{\mathrm{DM}}$ is the number density of DM, and $\sigma_{\mu,\mathrm{DM}}$ is the muon-DM cross section. For the case of ${\rm M}_{\mathrm{DM}}=500\,\mathrm{MeV}$, we simulate different signal MC samples with different $L$ assumption. Each signal MC sample include $10^8$ events. For $L=10^1\sim10^7$~km, it is expected to have $10^6\sim1$ events with a large scattering angle. The $\cos\theta$ distributions at the sea level are shown in Fig.~\ref{fig:costheta_signal_sea}. We can observe a significant discrepancy between the background distribution and signal distributions with $L=10$~km and 100~km. For $L\ge 100$~km, no significant differences are found. According to the $\chi^2$ test method, we calculate the $\chi^2/\mathrm{ndf}$ between the signal histogram and background histogram. The resulting $p$-values indicate that $L=10$~km ($\sigma_{\mu,\mathrm{DM}}=1.67\times10^{-6}$~cm$^{2}$) and 100~km ($1.67\times10^{-7}$~cm$^{2}$) signal assumptions can be rejected by more than $20\sigma$, while the significances for $L\ge 100$~km signal assumptions are lower then $5\sigma$. It indicates that the rejection ability for $M=500$~MeV DM is about $\sigma_{\mu,\mathrm{DM}}=10^{-7}\sim10^{-8}$~cm$^{2}$ for 1~m$^{2}$ detectors placed at a mountain with a latitude of 100~m and the sea level after one-year data taking.

\section{Next: Muon on target} 

Lepton beams may play an important role in the detection of light DM particles.
There are some proposals using electron beams for direct detection of sub-GeV DM particles~\cite{Essig:2012yx,Battaglieri:2020lds,Berlin:2020uwy,LDMX:2018cma,DarkSide:2022knj}. These approaches utilize electron beams to probe invisibly-decaying particles which couple to electrons. Some theoretical models like Muonphilic DM and Lepton portal DM~\cite{AlAli:2021let,Bai:2014osa} show that DM states only couple to the charged leptons, and predominantly interact with muons with a new type of interaction called muonic force. The quest for explaining the pronounced muon magnetic moment anomaly from the Fermilab Muon g-2 collaboration with 4.2 $\sigma$ discrepancy~\cite{Muong-2:2021ojo} drives people to search such DM preferentially interacting with muons through proton beam-dump experiments utilizing muon beams~\cite{Forbes:2022bvo}. The muonphilic DM scenario may reconcile this anomaly by introducing a force carrier particle, muonphilic bosons~\cite{Gninenko:2014pea}.
 
Accelerator muon beams have been proposed as a potential tool for detecting DM particles by NA64-$\mu$ experiment~\cite{Chen:2018vkr} (with first results public recently in Ref.~\cite{Andreev:2024sgn}) as well as ${\rm M}^3$~\cite{Kahn:2018cqs} and FNAL-$\mu$ experiment at Fermilab~\cite{Chen:2017awl}, and also for the future muon collider~\cite{Cesarotti:2022ttv}. The idea behind is to look for the muon deflection caused by scattering with DM as well as the the energy loss pattern of muons.  

Accordingly, in extension of above mentioned programs in previous sections by exploiting muon beams, we are aiming at further launching a muon missing momentum experiment in the future, which needs further optimization R\&D studies, especially when interfacing with domestic muon beams with lower energy. More details can be found in a forthcoming study by the authors.

\section{Summary and outlook} 
We propose here a set of new methods to directly detect light mass DM through its scattering with abundant atmospheric muons or accelerator beams. Firstly, we plan to use the free cosmic-ray muons interacting with dark matter in a volume surrounded by tracking detectors, to trace possible interaction between dark matter and muons. Secondly, we will interface our device with domestic or international muon beams. Due to much larger muon intensity and focused beam, we anticipate the detector can be made further compact and the resulting sensitivity on dark matter searches will be improved. In line with above projects, we will develop muon tomography methods and apply them on atmospheric and environmental sciences, archaeology and civil engineering. Furthermore, we will measure precisely directional distributions of cosmic-ray muons, either at mountain or sea level, and the differences may reveal possible information of dark matter distributed near the earth. In the future, we may also extend our study to muon on target experiments. 

Above various methods using either cosmic muons or accelerator muon beams, will provide a first direct probe on muon and dark matter interactions, with sensitivities spanning over different orders of magnitudes. Specifically, our methods can have advantages over `exotic' dark matters which are either muon-philic or slowed down due to some mechanism. In such a scenario, due to enlarged dark matter number density, sensitivity on dark matter and muon scattering cross section can reach near microbarn level. 

Over all, we believe all these searches complement and enrich current dark matter programs, and may stimulate further efforts.

\begin{acknowledgments}
This work is supported in part by the National Natural Science Foundation of China under Grants No. 12150005, No. 12325504, No. 12075004, and No. 12061141002.

Xudong Yu, Chen Zhou, Qite Li and Qiang Li contributed equally to this work.
%https://journals.aps.org/prl/authors
\end{acknowledgments}

\end{document}